\documentclass[twocolumn, showpacs, preprintnumbers]{revtex4}

\usepackage[dvips]{graphicx}
\usepackage{amsmath, amssymb}
\usepackage{dcolumn}
\usepackage{bm}
\usepackage{color}

\begin{document}

\title{Quasiclassical analysis of vortex lattice states in Rashba noncentrosymmetric superconductors}
\author{Yuichiro Dan and Ryusuke Ikeda}
\affiliation{Department of Physics, Kyoto University, Kyoto 606-8502, Japan}
\date{\today}

\begin{abstract}
Vortex lattice states occurring in noncentrosymmetric superconductors with a spin-orbit coupling of Rashba type under a magnetic field parallel to the symmetry plane are examined by assuming the $s$-wave pairing case and in an approach combining the quasiclassical theory with the Landau level expansion of the superconducting order parameter.
The resulting field-temperature phase diagrams include not only a discontinuous transition but a continuous crossover between different vortex lattice structures, and, further, a critical end point of a structural transition line is found at an intermediate field and a low temperature in the present approach.
It is pointed out that the strange field dependence of the vortex lattice structure is a consequence of that of its anisotropy stemming from the Rashba spin-orbit coupling, and that the critical end point is related to the helical phase modulation peculiar to these materials in the ideal Pauli-limited case.
Furthermore, calculation results on the local density of states detectable in STM experiments are also presented.
\end{abstract}


\maketitle

\section{Introduction}

Motivated by the recent revival of spatially modulated superconducting (SC) states induced by paramagnetic pair breaking (PPB) \cite{AI03,Kaur1,crosscriss,Voron}, noncentrosymmetric superconductors \cite{Baur} in nonzero magnetic fields have been studied in recent years as a novel type of system with a peculiar modulated SC state. 
In noncentrosymmetric superconductors, the lack of spatial inversion symmetry results in a splitting of the original Fermi surface into two sheets and makes effects of PPB anisotropic.
Then, this anisotropic PPB effect tends to create a helical modulation of the phase of the SC order parameter just in a specific direction \cite{Kaur1}.
In particular, it is remarkable that such a modulated state may be realized even in a small enough magnetic field \cite{Kaur1}, in contrast to the conventional Fulde-Ferrell-Larkin-Ovchinnikov (FFLO) states \cite{FFLO}, which do not appear unless the applied magnetic field reaches a high value of the order of the PPB field $H_{\rm P}$ at zero temperature.

In any type-II superconductors, however, when the applied magnetic field is higher than the so-called lower critical field $H_{c1}$, the field-induced vortices enter the SC material, and, if the PPB-induced helical direction is perpendicular to the applied field, the induced phase modulation may be absorbed in a nontrivial manner into a change of the vortex lattice pattern of the SC order parameter.
In fact, it has been pointed out in the Ginzburg-Landau (GL) approach that, in the case of noncentrosymmetric superconductors with an antisymmetric spin-orbit coupling of Rashba type \cite{Rashba}, any periodic phase modulation perpendicular to the field is gauged away in the order parameter solution so that the PPB-induced helical modulation cannot be seen in {\it bulk} Rashba superconductors.
However, it is unclear whether any effect of the helical modulation in the vortex-free limit does not occur even beyond the GL theory. 

In this paper, vortex lattice states in Rashba superconductors which occur when the magnetic field is parallel to the basal plane corresponding to the symmetry plane for the spatial inversion are examined beyond the GL approach and by combining the quasiclassical approach \cite{Eilenberger1, LO69}, which is widely exploited in microscopic analysis of superconductors \cite{Ichioka1, Nagai1, Vorontsov1} including that of multiband ones \cite{Kusunose1, Vorontsov2}, with the Landau level (LL) expansion of the order parameter \cite{Adachi1}.
It has been found in the previous GL approach \cite{Matsu,Hiasa1} that the structural symmetry of the vortex lattice in Rashba superconductors in the in-plane field configuration dramatically changes as the field increases through first-order transitions or continuous crossovers.
This is a consequence of the enhanced role of the higher LLs induced by the PPB.
It is difficult to describe such a field-dependent structural change of the vortex lattice in terms of the conventional method based on comparison in energy among a couple of {\it assumed} lattice structures.
On the other hand, the LL expansion has been regarded as a convenient tool in the GL approach which is not applicable to lower temperatures and lower fields.
However, the LL expansion of the order parameter has been applied to the quasiclassical (Eilenberger-Larkin-Ovchinnnikov) approach to examine the diamagnetic properties \cite{Adachi1} by incorporating an approximation analogous to the so-called Pesch approximation \cite{Pesch}.
We have chosen to use this LL expansion method in the quasiclassical approach \cite{Adachi1} to address the low-temperature vortex lattice states which cannot be examined in the GL approach \cite{Hiasa1}.

One of the main results in the present work is the presence of a critical end point of a first-order structural transition of the vortex lattice in the low-temperature and intermediate-field regime which cannot be well described by the previous GL approach \cite{Hiasa1}.
We argue that the presence of this critical end point is related to the helical phase modulation in the vortex-free limit mentioned above and to the field-induced compression of the vortex lattice structure due to the PPB.
Furthermore, as an electronic measure of the structure of the vortex lattice at each field and temperature, we calculate the local density of states (LDOS) in each vortex lattice.  

The rest of this paper is organized as follows.
In Sec.\,II, the electronic model examined in the present work and the content of our theoretical approach are explained.
The obtained phase diagrams and the vortex lattice structures are shown and discussed in Sec.\,III together with the calculation results on the LDOS.
In Sec.\,IV, our results are summarized, and the details of the quasiclassical analysis we have used are explained in Appendices.

\section{Model and Theoretical Approach} \label{sec:theory}

\begin{figure}
\centering
\includegraphics[scale=0.5]{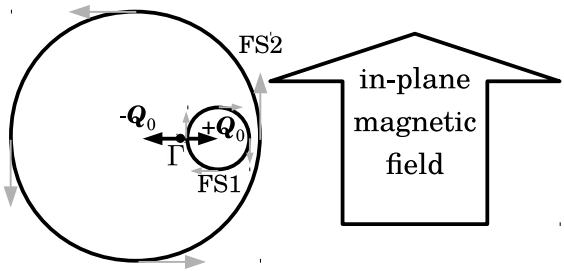}
\caption{Fermi surfaces (FS1 and FS2) under an in-plane magnetic field.
The gray arrows indicate the direction of the spin fixed by the spin-orbit coupling of Rashba type.
Each surface shifts oppositely from $\Gamma$ by $\pm\bm{Q}_0$.
The vector $\bm{Q}_0$ is defined by Eq.\,(\ref{eq:def_Q0_main}) and in Appendix A.}
\label{fig:FS}
\end{figure}

Following the previous work \cite{Matsu}, we start from the Hamiltonian with only an $s$-wave attractive interaction: 
\begin{gather}
\mathcal{H} = \mathcal{H}_{\mathrm{single}} + \mathcal{H}_{\mathrm{int}}, \\
\mathcal{H}_{\mathrm{single}} = \sum_{\bm{k},\alpha,\beta}c^{\dag}_{\bm{k}\alpha}[\varepsilon_{\bm{k}}\sigma_0 + \zeta \bm{g}_{\bm{k}}\cdot\bm{\sigma}]_{\alpha,\beta}c_{\bm{k}\beta} \notag\\
+ \int d^3\bm{r}\sum_{\alpha\beta}c^{\dag}_{\alpha}(\bm{r})\mu_s\bm{B}(\bm{r})\cdot\bm{\sigma}_{\alpha,\beta}c_{\beta}(\bm{r}),\label{eq:noninteracting_hamiltonian}\\
\mathcal{H}_{\mathrm{int}} = -\frac{gV}{4}\sum_{\bm{q}}\Psi_{\bm{q}}^{\dag}\Psi_{\bm{q}},
\end{gather}
where $c_{\bm{k}\alpha}$ is the annihilation operator of an electron with momentum $\bm{k}$ and spin $\alpha (=\uparrow,\downarrow)$, $c_{\alpha}(\bm{r})$ is that at position $\bm{r}$, the $\sigma_{\mu}$'s ($\mu=0,1,2,3$) are the Pauli matrices, $g$ is the coupling constant,  $V$ is the volume of the system,
\begin{equation}
\Psi_{\bm{q}}=\frac{1}{V}\sum_{\bm{k},\alpha,\beta}c_{-\bm{k}+\bm{q}/2,\alpha}(-i\sigma_2)_{\alpha\beta}c_{\bm{k}+\bm{q}/2,\beta}
\end{equation}
is the field operator of a spin-singlet $s$-wave Cooper pair with the total momentum $\bm{q}$, 
$\mathcal{H}_{\mathrm{single}}$ is the noninteracting term of the quasiparticle Hamiltonian, and $\mathcal{H}_{\mathrm{int}}$ is the $s$-wave pairing interaction term. Regarding the centrosymmetric term of the quasiparticle dispersion 
$\varepsilon_{\bm{k}}$, we assume the quasi-two-dimensional form 
\begin{equation}
\varepsilon_{\bm{k}} = \frac{1}{2m}(k_x^2 + k_y^2) + J(1-\cos{k_zd}),
\end{equation}
where $m$ is the effective mass of a quasiparticle, and $d$ is the lattice constant in the $c$-axis direction.
The antisymmetric spin-orbit coupling (ASOC) of Rashba type is described by 
\begin{equation}
\bm{g}(\bm{k}) = \frac{\bm{k}_\perp \times\hat{\bm{z}}}{k_{\mathrm{F}}}, 
\end{equation}
where $\bm{k}_\perp=\bm{k} - k_z \hat{\bm{z}}$ is the two-dimensional wave vector, $k_{\mathrm{F}} = \sqrt{2mE_{\mathrm{F}}}$, $E_{\rm F}$ is the bare Fermi energy, $\hat{\bm{z}}$ is the unit vector in the direction of the broken inversion symmetry, and 
$\zeta$ is the strength of the ASOC. 
Throughout this paper, the $xy$ plane is the basal plane on the broken inversion symmetry, and $J$ is the interplane coupling constant.
Then, the anisotropy of the coherent lengths is given by 
\begin{equation}
\gamma = \frac{\xi_x}{\xi_z} = \sqrt{\frac{\left<v_{x}^2\right>_{\mathrm{FS}}}{\left<v_{z}^2\right>_{\mathrm{FS}}}} = \frac{2\sqrt{1 - J/E_{\mathrm{F}}}}{k_{\mathrm{F}}d J/E_{\mathrm{F}}},
\end{equation}
where $\xi_x$ and $\xi_z$ are the in-plane and the out-of-plane coherent length, respectively. The angle average over the Fermi surface is defined as 
\begin{equation}
\left<h(\bm{k})\right>_{\mathrm{FS}} = \int_{-\pi/d}^{\pi/d}\frac{dk_z}{2\pi/d}\int_{0}^{2\pi}\frac{d\phi_{\bm{k}}}{2\pi}h(\bm{k})
\end{equation}
for an arbitrary function $h(\bm{k})$, where $\phi_{\bm{k}} = \mathrm{tan}^{-1}\frac{k_y}{k_x}$.
In addition, $\mu_s$ is the magnetic moment of the spin of a quasiparticle, and $\bm{B}$ is the magnetic flux density.
Although in model (1) the orbital effect of a magnetic field is not incorporated, it can be readily included through the Peierls substitution:
\begin{equation}
\bm{k} \rightarrow \bm{k} + e\bm{A},
\end{equation}
where $-e$ is the electronic charge, and $\bm{A}$ is the vector potential associated with $\bm{B}$. 

As in the previous works \cite{Matsu,Hiasa1,Kaur1}, we focus on the case with such a realistically large ASOC that 
\begin{equation}
{\rm Max}(T_c, \,\,\, \mu_s |\bm{B}|) \ll |\zeta| \ll E_{\rm F},
\label{inequal}
\end{equation}
where $T_c$ is the transition temperature at zero field.
The smallness of the ratio $\mathrm{Max}(T_c, \mu_s |\bm{B}|)/|\zeta|$ results in simplifying the mean-field (BCS) Hamiltonian under which the Eilenberger equations are constructed.
Before constructing the mean-field quasiparticle Hamiltonian, however, the quadratic term ${\cal H}_{\rm single}$ has to be diagonalized.
After diagonalization, we encounter a quasiparticle Hamiltonian consisting of two independent bands.
In Fig.\,\ref{fig:FS}, the resulting Fermi surfaces are sketched.
On the other hand, this diagonalization induces pairing interactions between the split two bands.
However, these interband interaction terms are relatively of $O((\mathrm{Max}(T_c, \mu_s |\bm{B}|)/\zeta)^3)$ and hence, can be neglected.

Then, as explained in Appendix B, the corresponding transformation of the Green's functions leads to the Gor'kov equations consisting only of intraband terms.
As a result, we obtain the following Eilenberger equations:
\begin{eqnarray}
\left[2 \left\{ \omega_n + i(-1)^{a+1}\mu_s\hat{\bm{g}}_{\bm{k}}\cdot\bm{B}\right\} + i\bm{v}_{\mathrm{F}}\cdot\bm{\Pi}\right] f_a &\notag\\
&\!\!\!\!\!\!\!\!\!\!\!\!\!\!\!\!\!\!\!\!\!\!\!\!\!= -2 i w_a \Delta g_a, \label{eq:eilenberger_1}
\end{eqnarray}
\begin{eqnarray}
\left[2 \left\{ \omega_n + i(-1)^{a+1}\mu_s\hat{\bm{g}}_{\bm{k}}\cdot\bm{B}\right\} + i\bm{v}_{\mathrm{F}}\cdot\bm{\Pi}^{\ast}\right] \bar{f}_a &\notag\\
&\!\!\!\!\!\!\!\!\!\!\!\!\!\!\!\!\!\!\!\!\!\!\!\!\!= - 2 i w_a^* \Delta^* g_a, \label{eq:eilenberger_2}
\end{eqnarray}
\begin{equation}
g_a = - \sqrt{1 + f_a \bar{f}_a} \, \, \, (\mathrm{Re}\,g_a < 0), 
\end{equation}
The indices $a(=1,2)$ specify the two split bands.
Furthermore, $g_a$, $f_a$, and $\bar{f}_a$ are the normal and anomalous quasiclassical Green's functions, $\bm{v}_{\mathrm{F}}$ is the Fermi velocity on each FS which has the same value for both FSs up to the lowest order in $\zeta/E_{\mathrm{F}}$ and $J/E_{\mathrm{F}}$ (see Ref.\,\cite{Hayashi1, Yip1} and also Appendix E in the present work), 
\begin{align}
w_a =& (-1)^ai\exp(i (-1)^a \phi_{\bm{k}}) \notag\\
=& (-1)^ai|{\bf k}_\perp|^{-1} (k_x + i (-1)^a k_y)
\end{align}
is a pairing function associated with the two bands occurring after the diagonalization, 
\begin{eqnarray}
\bm{\Pi} =& -i\nabla +2e\bm{A},\notag\\
\bm{\Pi}^{\ast} =& i\nabla + 2e\bm{A},
\end{eqnarray}
and $\omega_n (>0)$ is the fermion Matsubara frequency. This set of equations is equivalent to that used in previous works \cite{Hayashi1, Hayashi2, Hayashi3} except for the inclusion of the Zeeman effect. 
Note that the two bands split by the ASOC are coupled with each other only through the shared order parameter $\Delta$ due to the condition $\zeta\gg T_c$.

Next, to solve Eqs.\,(\ref{eq:eilenberger_1}) and (\ref{eq:eilenberger_2}), we assume the type-II limit hereafter so that $\bm{B} = \bm{H}$, where $\bm{H}$ is the applied field along the $y$ axis.
In addition, following Adachi \textit{et al}. \cite{Adachi1}, the Landau level (LL) expansion of $\Delta$ is used in the quasiclassical approach. 
This is because the conventional treatment based on comparison in the free energy among a couple of {\it assumed} structures is not fruitful in the present issue where field-dependent and continuous changes of the vortex lattice structure are expected to occur \cite{Matsu,Hiasa1}.
Nevertheless, it is difficult to find an exact solution of Eqs.\,(\ref{eq:eilenberger_1}) and (\ref{eq:eilenberger_2}) based on the LL expansion method, and hence, we adopt an approximation \cite{Adachi1} underestimating spatial variations of $|\Delta|^2$ to be included in the normal Green's function $g_a$ corresponding to an analog of the Pesch approximation \cite{Pesch}.
The result following from this approximation, dubbed the ``approximate solution'' in Ref.\,\cite{Adachi1}, was argued there to be valid  not only near the $H_{c2}$ line but also in the low-field region as long as thermodynamic quantities are considered \cite{Adachi2}.
Since the central part of our present work is to find the vortex lattice structure with the lowest free energy at each field and temperature, this approximated method can be used to determine the lattice shape to be realized over wide field and temperature ranges in the phase diagram.
Nevertheless, one should keep in mind that, for the purpose of resolving a fine spatial structure, e.g., a single vortex core structure, this approximation gets less precise at lower fields \cite{Adachi1}.

Then, Eqs.\,(\ref{eq:eilenberger_1}) and (\ref{eq:eilenberger_2}) are rewritten as 
\begin{equation}
f_a = g_a\Phi_a,\ \ \bar{f_a} = g_a\bar{\Phi}_a,\ \ g_a = -1/\sqrt{1 - \Phi_a\bar{\Phi}_a},
\end{equation}
where
\begin{eqnarray}
\Phi_a &=& -2iw_a\left[2 \left\{ \omega_n + i(-1)^{a+1}\mu_s\hat{\bm{g}}_{\bm{k}}\cdot\bm{H}\right\} + i\bm{v}_{\mathrm{F}}\cdot\bm{\Pi}\right]^{-1}\Delta, \notag \\
\bar{\Phi}_a &=& -2iw_a^{\ast}\left[2 \left\{ \omega_n + i(-1)^{a+1}\mu_s\hat{\bm{g}}_{\bm{k}}\cdot\bm{H}\right\} + i\bm{v}_{\mathrm{F}}\cdot\bm{\Pi}^{\ast}\right]^{-1}\Delta^{\ast}.\notag\\
\end{eqnarray}
The order parameter $\Delta$ can be expanded in terms of LLs:
\begin{equation}
\Delta(\bm{r}) = \sum_{N} d_N \psi_N(\bm{r}), 
\end{equation}
where
\begin{eqnarray}
\psi_N(\bm{r}) &=& e^{-\bm{Q}\cdot\bm{r}}\sqrt{\nu}\sum_{m\in\mathbb{Z}}e^{-i\pi\lambda m^2}e^{im\nu \gamma^{1/2}z/r_H} \nonumber \\
&\times& \Psi_N(\gamma^{-1/2}x/r_H - m\nu)\label{eq:LL}
\end{eqnarray}
is the $N$th LL ($N \geq 0$) when the Landau gauge $\bm{A} = -Hx\hat{\bm{z}}$ is chosen. 
Here, $r_H = 1/\sqrt{2eH}$ is the magnetic length which characterizes the spacing between two vortices. 

The wave vector of the helical phase $\bm{Q}$ is nonvanishing as far as $\delta N$ is finite.
Throughout the present work, $\bm{Q}$ is fixed to $2\delta N\bm{Q}_0$, where
\begin{equation}
\bm{Q}_0 = \frac{\mu_s H}{v_{\mathrm{F}}}\hat{\bm{x}}\label{eq:def_Q0_main}
\end{equation}
is the shift of the Fermi surfaces (see Fig.\,\ref{fig:FS} and Appendix A), and the deviation of the true $\bm{Q}$ from $2 \delta N \bm{Q}_0$ is assumed to be compensated by incorporating as many LLs as possible.
It is originally known that the identification $\bm{Q}=2 \delta N \bm{Q}_0$ is safely valid near $H_{c2}(T)$ at high temperatures \cite{Kaur1,Matsu} where the higher gradients may be neglected.
In the GL approach in Ref.\,\cite{Hiasa1}, the validity of this identification has been tested in the simplest $s$-wave pairing case by comparing with the exact result obtained by determining the $\bm{Q}$ value minimizing the free energy at each magnetic field, and the simplified treatment based on the identification $\bm{Q}=2 \delta N \bm{Q}_0$ has been shown not to affect the phase diagram even quantitatively (see Fig.\,3 and its related discussions in Ref.\,\cite{Hiasa1}). 

\begin{figure}[b]
\centering
\includegraphics[scale=0.5]{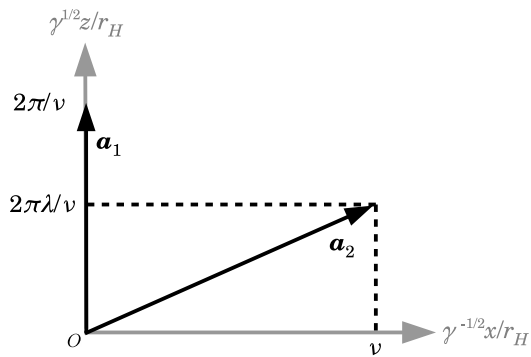}
\caption{Definition of the parameters $\nu$ and $\lambda$, which represents the shape of the vortex lattice. Here, $\bm{a}_1$ and $\bm{a}_2$ are principal lattice vectors of the lattice.}
\label{fig:lattice_parameters}
\end{figure}

The function $\Psi_N$ is expressed by the $N$th Hermite polynomial $H_N$ as follows:
\begin{equation}
\Psi_N(x) = \frac{H_N(x)e^{-x^2/2}}{\sqrt{2^NN!}\pi^{1/4}}.
\end{equation}
Parameters $\nu$ and $\lambda$, which represent the lattice shape, are defined as in Fig.\,\ref{fig:lattice_parameters}. 
Due to this expansion, the function $\Phi_a$ can be described as the linear algebraic expression:
\begin{equation}
\Phi_a = -2iw_a \sum_{M,N}\psi_M\mathcal{M}^a_{MN}d_N\label{eq:Phi_lin_alg},
\end{equation}
and $\bar{\Phi}_a$ can be calculated from the relation,
\begin{equation}
\bar{\Phi}_a(\bm{k}) = \Phi_a(-\bm{k})^{\ast}, \label{eq:phi_symmetry}
\end{equation}
which can be proved with the symmetry relations presented in Appendix D.
The matrix $\mathcal{M}^a_{MN}$ is defined as follows:
\begin{align}
\mathcal{M}^a_{MN} &=\int_{0}^{\infty}d\rho e^{-(2\omega_n\rho + |s|^2\rho^2/2)}\notag\\
&\times e^{i\left\{\bm{v}_{\mathrm{F}}\cdot\bm{Q}-2(-1)^{a+1}\mu_s\hat{\bm{g}}_{\bm{k}}\cdot\bm{H}\right\}\rho}\mathcal{L}_{MN}(-is^{\ast}\rho),
\end{align}
which is remarkably independent of the lattice shape, where
\begin{gather}
s = \frac{\gamma^{1/2}v_{\mathrm{F},z} + i\gamma^{-1/2}v_{\mathrm{F},x}}{\sqrt{2r_H^2}},\\
\mathcal{L}_{MN}(z) = \sum_{l=0}^{\min(M,N)}\frac{\sqrt{M!N!}}{(M-l)!(N-l)!l!}(z)^{M-l}(-z^{\ast})^{N-l}.
\end{gather}
The derivation of Eq.\,(\ref{eq:Phi_lin_alg}) is shown in Appendix G. 

Since the shape of the vortex lattice to be realized is determined through minimization of the free energy per unit volume, we need an expression of the free energy represented by the quasiclassical Green's functions.
According to the theory of Eilenberger \cite{Eilenberger1}, the free energy $F$ is calculated through the variational principle in the form
\begin{widetext}
\begin{equation}
F = N\int d^3\bm{r}\left[ |\Delta|^2\ln{\frac{T}{T_\mathrm{c}}} + 2\pi T\sum_{\omega_n>0}\sum_{a=1,2}\frac{1 + (-1)^a\delta N}{2}\left<\left(\frac{|\Delta|^2}{\omega_n} - \frac{iw_a\Delta\bar{f}_a + i\Delta^{\ast}w_a^{\ast}f_a}{g_a-1}\right)\right>_{\mathrm{FS}} \right], 
\label{eq:free_energy}
\end{equation}
\end{widetext}
where $T$ is the temperature,
\begin{equation}
N = \frac{N_1 + N_2}{2},
\end{equation}
\begin{equation}
\delta N = \frac{N_2- N_1}{N_1 + N_2} \left( \propto \frac{\zeta}{E_{\mathrm{F}}} \right)
\end{equation}
and $N_a$ is the normal DOS on the $a$th FS. 
More details on the derivation of Eq.\,(\ref{eq:free_energy}) are given in Appendix F.

To make the optimization feasible, the LLs are transformed into the linear combinations of the LLs by diagonalizing the quadratic term, $F_2$, of the free energy with respect to the order parameter $\Delta$. Using the parameter integral $A^{-1} =  \int_0^\infty d\rho \exp(-\rho A)$, the quadratic term $F_2$ is easily obtained from Eq.\,(\ref{eq:free_energy}) and becomes 
\begin{widetext}
\begin{align}
\frac{F_2}{V} &= N\sum_{M,N}d_M\left[\delta_{M,N}\ln\frac{T}{T_c} + \int_{0}^{\infty}d\rho f(\rho)\left\{ \delta_{M,N} \right.\right.\notag\\
&\left.\left. -\sum_a\frac{1+(-1)^a\delta N}{2}\left< |w_a|^2e^{-|s|^2\rho^2/2}\mathcal{L}_{M,N}(-is^{\ast}\rho)e^{i\left\{\bm{v}_{\mathrm{F}}\cdot\bm{Q}-2(-1)^{a+1}\mu_s\hat{\bm{g}}_{\bm{k}}\cdot\bm{H}\right\}\rho} \right>_{\mathrm{FS}} \right\}\right]d_N \label{eq:free_energy_second_order_term}
\end{align}
\end{widetext}
which coincides with the expression obtained in the previous study \cite{Hiasa1}, where
\begin{equation}
f(\rho) = \frac{2\pi T}{\sinh(2\pi T\rho)}.
\end{equation}
The matrix to be diagonalized is the expression between the square brackets in Eq.\,(\ref{eq:free_energy_second_order_term}), and the resulting modes are the linear combinations of the LLs.
If the modes resulting from the diagonalization are separated in energy from one another, we only have to select just the mode with the lowest energy to obtain the vortex lattice structure in equilibrium, because an energy difference between lattice structures is usually much smaller \cite{Lasher}. 
Thus, we have three variational parameters: $\nu$, $\lambda$, and the amplitude of the relevant mode.

After the diagonalization, the $H_{c2}$ line is determined, as usual, as the line in the $H$-$T$ phase diagram on which the eigenvalue of the lowest energy mode changes its sign upon cooling.
As already noted elsewhere \cite{Matsu,Hiasa1},  the transition at the $H_{c2}(T)$ line defined in the mean-field approximation is, in contrast to that in the centrosymmetric case \cite{AI03}, of second order irrespective of the temperature and the strength of the PPB effect.
Indeed, we have confirmed that the quartic term in $F$ with respect to $\Delta$ is positive for the mode with the lowest eigenvalue even in the high-field and low-temperature region when $\delta N = 0$, where the suppression of the Zeeman effect due to the ASOC is the weakest. 

Furthermore, as a physical quantity testable in STM experiments and reflecting the vortex lattice structures, we have considered the LDOS.
To obtain this quantity, the analytic continuation, $i\omega_n \rightarrow E + i\eta$ is performed, where $\eta$ is an infinitesimal and positive.
In the present formalism, this is equivalent to the replacement $i\omega_n \rightarrow E + i\eta$ in $\mathcal{M}^a_{MN}$, which leads instantly to the retarded quasiclassical Green's function $g_a^{\mathrm{R}}$.
Then, we have 
\begin{equation}
N(\bm{r};E) = -N\sum_{a}(1 + (-1)^a\delta N)\mathrm{Re}\,\left<g_a^{\mathrm{R}}(\bm{r}; \hat{\bm{k}}, E)\right>_{\mathrm{FS}}\label{eq:LDOS}
\end{equation}
as the LDOS (see Appendix H). 
In this case, note that the relation between the retarded versions of $\Phi_a$ and $\bar{\Phi}_a$ is given not by Eq.\,(\ref{eq:phi_symmetry}) but by 
\begin{equation}
\bar{\Phi}_a(\bm{k}, E) = \Phi_a(-\bm{k}, -E)^{\ast} \label{eq:phi_symmetry_retarded}.
\end{equation}

\section{Results}

In this section, our calculation results on the phase diagram are shown and explained. 
In all of our calculation results presented in this paper, we have commonly used the parameter values $J/E_{\mathrm{F}}=0.1$, $H_{\mathrm{orb}}^{\mathrm{2D}}/H_{\mathrm{P}}(\propto \mu_s) = 2.0$, and $d = \pi/k_{\mathrm{F}}$, which are the same values as those in Fig.\,3 in Ref.\,\cite{Hiasa1}, where $H_{\mathrm{orb}}^{\mathrm{2D}}$ ($= 0.56\phi_0/2\pi\xi_0^2$) and $H_{\mathrm{P}}$ ($= 1.25T_c/\mu_s$) are the orbital pair breaking field in 2D systems and the paramagnetic pair breaking field at zero temperature, respectively, $\phi_0 = \pi/e$ is the flux quantum, and $\xi_0 = v_F/2\pi T_c$ is the coherent length in the directions parallel to the basal plane. 

\subsection{$\delta N=0$ case}

In this subsection, the resulting phase diagram in the limiting case with $\delta N=0$ is explained.
According to the inequality (\ref{inequal}), this case corresponds to the limit of a large bandwidth. 

\begin{figure}
\centering
\includegraphics[scale=0.6]{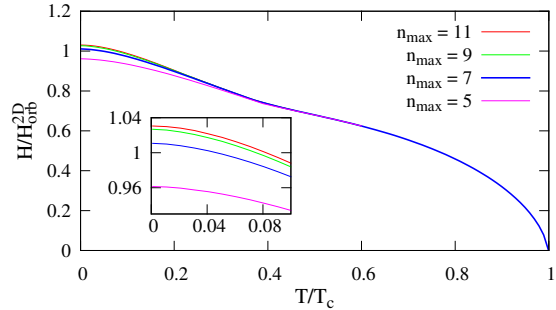}
\caption{(Color online) Dependence of the $H_{c2}$ curve on the number of LLs incorporated in the calculation in the $\delta N=0$ case, where $n_{\mathrm{max}}$ is the index of the highest LL incorporated.}
\label{fig:hc2_compare_deltaN0}
\end{figure}

First, the number of the LLs to be incorporated in our calculation should be determined.
As more LLs are included, the resulting $H_{c2}$ value at each temperature becomes higher.
In Fig.\,\ref{fig:hc2_compare_deltaN0}, such an example of the dependence of $H_{c2}(T)$ on the number of the incorporated LLs is presented, where $n_{\rm max}$ is the index of the highest LL incorporated.
Ideally, the saturation of $H_{c2}$ value should be reached by a finite value of $n_{\rm max}$.
Based on the $n_{\rm max}$ dependence of the $H_{c2}(T)$ curve obtained in Fig.\,\ref{fig:hc2_compare_deltaN0}, we have kept just the lowest eight LLs to determine the vortex lattice structure, as in Ref.\,\cite{Hiasa1}. 

\begin{figure}
\centering
\includegraphics[scale=0.6]{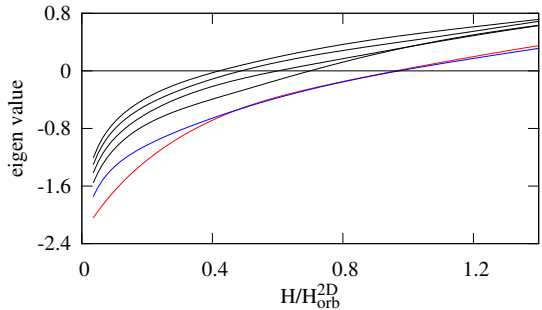}
\caption{(Color online) Field dependence of eigenvalues of the modes obtained by diagonalizing $F_2$ at $T = 0.1 T_c$ when $\delta N = 0$.
The modes with the lowest two eigenvalues (the red and blue lines) are nearly degenerate for $H > 0.4 H_{\mathrm{orb}}^{\mathrm{2D}}$ with each other and cross at $H = 0.48 H_{\mathrm{orb}}^{\mathrm{2D}}$.}
\label{fig:mode_T1_deltaN0}
\end{figure}

Next, the details of the mode splittings resulting from the diagonalization in $F_2$ are explained.
In Fig.\,\ref{fig:mode_T1_deltaN0}, the field dependencies of the eigenvalues of the diagonalized modes are shown at a low temperature.
In this $\delta N=0$ case, the lowest two modes are found to be nearly degenerate for $H > 0.4 H_{\mathrm{orb}}^{\mathrm{2D}}$.
Thus, the free energies resulting from the two modes have been calculated individually and compared with each other to determine the vortex lattice structure in equilibrium. 

\begin{figure}
\centering
\includegraphics[scale=0.6]{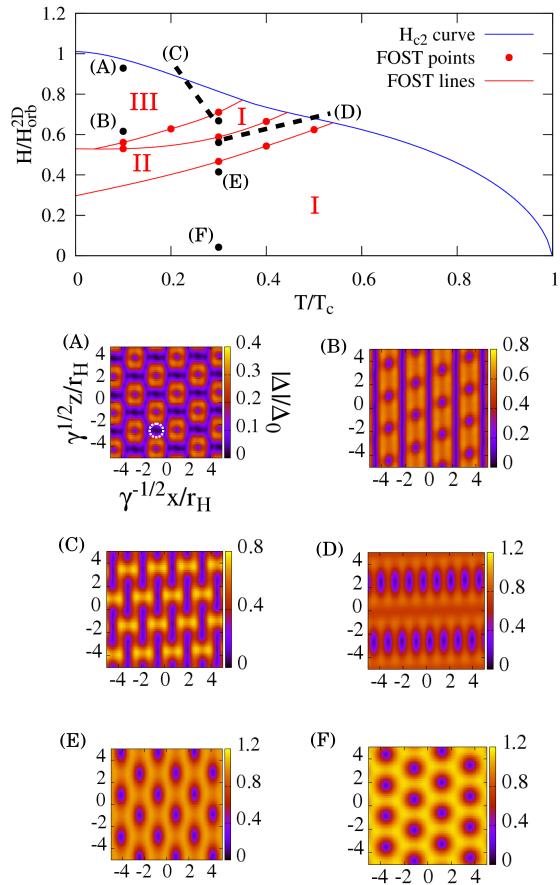}
\caption{(Color online) Resulting phase diagram and vortex lattice states appearing when $\delta N = 0$.
The first-order structural transition (FOST) points are numerically determined, and the line connecting between them is a guide to the eye. 
Phases I and II are stretched triangular lattice phases.
Phase III is a modulated triangular lattice phase.
(B) shows that the modulation is along the direction of the shift of the Fermi surfaces. The figures 
(C), (E), and (F) indicate that the lattice is compressed along the $x$ axis with increasing field. 
In (A), the spatial modulation in the region surrounded by the broken white circle does not imply the presence of an additional vortex there. 
Here, $\Delta_0$ is the magnitude of the order parameter at each temperature in the absence of magnetic fields. 
}
\label{fig:phase_diagram_deltaN0}
\end{figure}

The resulting phase diagram is shown in Fig.\,\ref{fig:phase_diagram_deltaN0}, in which there are three phases (I--III) separated by first-order structural transitions (FOSTs). 
The structure in phases I and II may be regarded as a stretched triangular lattice.
On the other hand, a {\it crossover} from a one-dimensional-like structure, in which a vortex layer and a nodal line are alternating, to a honeycomb-like vortex lattice occurs in phase III.
In the former (low-field) structure of phase III, the alternation occurs along the $x$ axis, namely, the direction of the shift of the Fermi surfaces. 

\begin{figure}
\centering
\includegraphics[scale=0.6]{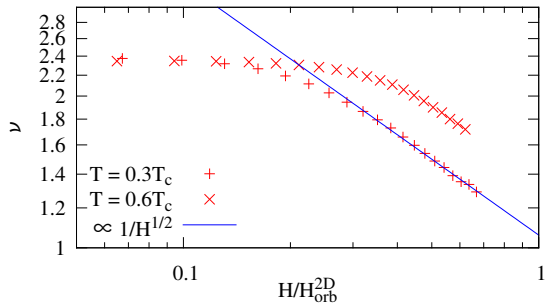}
\caption{(Color online) Field dependence of $\nu$, which is the lattice spacing along the $x$ axis (see Fig.\,\ref{fig:lattice_parameters}), of the lattice in phase I of Fig.\,\ref{fig:phase_diagram_deltaN0}.
Note that $\nu$ is almost proportional to $1/\sqrt{H}$ for $H > 0.3 H_{\mathrm{orb}}^{\mathrm{2D}}$ at $T=0.3T_c$.}
\label{fig:shape_compare_deltaN0}
\end{figure}

The most remarkable character seen commonly in these vortex states is the lattice compression parallel to the $x$ axis occurring with the field increasing. 
This is a consequence of the shift of Fermi surfaces caused by the in-plane applied field due to the interplay between the ASOC and the PPB effect.
Without PPB, no such field-induced anisotropy arises.
To avoid any confusion, the length scales in the $x$ and $z$ directions are measured in units of $r_H \gamma^{1/2}$ and $r_H \gamma^{-1/2}$, respectively, hereafter.
Reflecting the above-mentioned shift of the Fermi surfaces, the system favors the periodicity proportional to $1/(Q_0 r_H) \propto 1/H^{1/2}$ along the $x$ axis in real space.
On the other hand, since, due to the flux quantization, the area of the unit cell of the vortex lattice (in the above-mentioned units) is kept constant, the lattice spacing parallel to the $z$ axis is expanded with the field increasing. 
In this manner, the field-induced lattice compression in the $x$ direction is explained (see also Fig.\,\ref{fig:shape_compare_deltaN0}). 

This field-induced compression parallel to the $x$ axis tends to induce FOSTs between different vortex lattice symmetries in different ways.
In general, a compression in one direction merely enhances the anisotropy of the lattice, and a superfluous compression accompanied by no change of the lattice symmetry would lead to some energy cost.
Then, a FOST to a more isotropic lattice state may occur.
When a couple of lattice symmetries are competitive in energy to each other, however, it is possible for a FOST to occur between the two states without releasing the anisotropy.
In the present $\delta N=0$ case, the FOST of the latter type seems to be realized between the I and II phases in the low-field regime where the vortex lattice solution is formed in terms of only the LLs with even indices.
On the other hand, in higher fields, the vortex lattice consisting only of the LLs with odd indices is favored because of enhanced roles of the PPB in higher fields.
The above-mentioned release of the anisotropy is realized through the FOST between I and III phases together with this switching in the description of the order parameter from the even to odd LLs. 

On the other hand, at a glance, one might wonder why no FOST occurs between the structures (A) and (B) if noting the appearance in (A) of an additional modulation of the order parameter amplitude indicated by the broken white circle.
However, this modulation suggesting a node of the order parameter is not accompanied by any nonvanishing winding number and thus, is not a genuine vortex but just a modulated structure with a low but nonvanishing amplitude of the order parameter. 
Hence, the structure change between (A) and (B) can occur gradually and continuously to compensate the anisotropy of the vortex lattice as the field increases, which leads to the continuous crossover between them rather than a FOST. 

As mentioned above, the appearance of the vortex lattice consisting only of odd LLs in higher fields stems from the PPB effect, and consequently, the resulting vortex lattices in higher fields are mostly occupied by the spatial regions in which the order parameter amplitude $|\Delta|$ nearly vanishes.
To correctly describe such vortex lattices with PPB-induced additional modulations on the length scales of the magnetic length $r_H=\sqrt{\phi_0/(2 \pi H)}$, the nonlocality needs to be taken into account properly in the terms distinguishing different lattice structures in the free energy.
In the previous works based on the GL free energy kept up to the quartic order in the order parameter $\Delta$, the quartic term has been assumed in a spatially local form.
In describing details of the lattice structure, this local form is insufficient particularly in higher fields where the PPB effect is not negligible.
In fact, the resulting structures in the phase III are different from those in the previous GL approach. On the other hand, the field-induced transition from a triangular structure to another one, namely, from I to II in Fig.\,\ref{fig:phase_diagram_deltaN0}, is qualitatively similar to the previous one \cite{Matsu,Hiasa1}. 

\subsection{$\delta N=0.1$ case}

Next, we turn to a more realistic case with a nonvanishing $\delta N$ or a finite bandwidth.
The choice of the value $\delta N=0.1$ seems to be reasonable if one images the materials including $\mathrm{CePt_3Si}$ \cite{Samokhin1} as the corresponding model systems.

\begin{figure}
\centering
\includegraphics[scale=0.6]{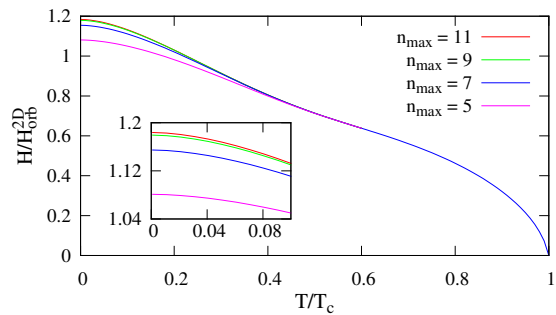}
\caption{(Color online) Dependence of the $H_{c2}$ curve on the number of LLs incorporated in calculation when $\delta N=0.1$. As in the $\delta N=0$ case, we have assumed $n_{\rm max}=7$.}
\label{fig:hc2_compare_deltaN1}
\end{figure}

\begin{figure}
\centering
\includegraphics[scale=0.6]{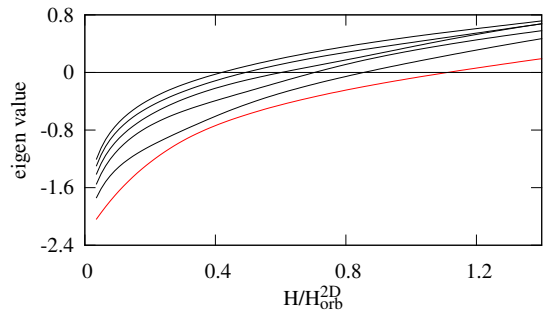}
\caption{(Color online) Field dependence of the eigenvalues of the modes obtained by diagonalizing $F_2$ at $T=0.1T_c$ when $\delta N = 0.1$.
The mode with the lowest eigenvalue (red line) is dominant at any field.}
\label{fig:mode_T1_deltaN1}
\end{figure}

As in the $\delta N=0$ case, the phase diagram is examined by including the lowest eight LLs (see Fig.\,\ref{fig:hc2_compare_deltaN1}). 
In this case, the even and odd LLs are mixed in every mode resulting from the diagonalization.
Furthermore, as seen in Fig.\,\ref{fig:mode_T1_deltaN1}, there is no competition between the modes, and we have a well-defined mode with the lowest energy eigenvalue. Thus, we only have to focus on this mode to determine the vortex lattice structure at each field and temperature based on the free energy (\ref{eq:free_energy}). 

\begin{figure}
\centering
\includegraphics[scale=0.6]{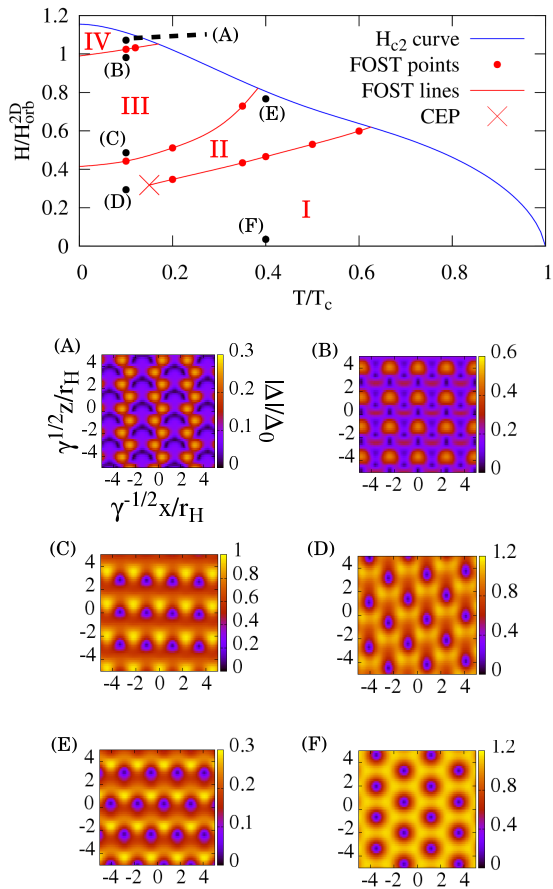}
\caption{
(Color online) Phase diagram and vortex lattice states when $\delta N = 0.1$.
The first order structural transition (FOST) points are numerically determined, and the line connecting between them is a guide to the eye.
Phases I and II are stretched triangular lattice phases, while phase III is a rectangular lattice one.
Furthermore, phase IV is a modulated triangular lattice phase.
Figure (A) shows that the modulation develops along the direction of the shift of the Fermi surfaces with the field increasing.
(D), (E) and (F) can be continuously transformed to one another circumventing the critical end point (CEP). 
}
\label{fig:phase_diagram_deltaN1}
\end{figure}

Figure\,\ref{fig:phase_diagram_deltaN1} shows the resultant phase diagram and vortex states. 
There, the phase I and II are stretched triangular lattice phases, while a rectangular lattice is stable in the phase III. 
The phase IV is characterized by a modulated triangular lattice structure. 
The most remarkable difference of this phase diagram from Fig.\,\ref{fig:phase_diagram_deltaN0} in the $\delta N=0$ case is the emergence of a critical end point of the FOST line between phases I and II. 

\begin{figure}
\centering
\includegraphics[scale=0.6]{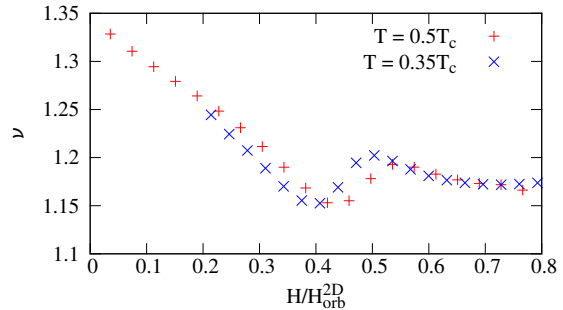}
\caption{
(Color online) Field dependence of $\nu$, the lattice spacing parallel to the $x$ axis, of the states in phase II of Fig.\,\ref{fig:phase_diagram_deltaN1}. 
The upturn of the width for $0.4 H_{\mathrm{orb}}^{\mathrm{2D}} < H < 0.5 H_{\mathrm{orb}}^{\mathrm{2D}}$ means that the field-induced compression parallel to the $x$ axis is weakened in higher fields. As for the definition of $\nu$, see Fig.\,\ref{fig:lattice_parameters}.}
\label{fig:shape_compare_deltaN1}
\end{figure}

First, to elucidate the effect of finite $\delta N$, as in the former $\delta N=0$ case, the relation between the lattice spacing along the $x$ axis and the applied field has been plotted.
It is remarkable in Fig.\,\ref{fig:shape_compare_deltaN1} that, although the spacing in the $x$ direction, broadly speaking, shrinks with the field increasing reflecting the compression induced by the ASOC, an upturn appears in its field dependence.
This seems to result from the appearance of another periodicity caused by the emergence of the helical phase in the vortex-free 
case \cite{Kaur1}. The wave vector $Q$ of the helical phase is known to be proportional to $\delta N$ (see the description below Eq.\,(\ref{eq:LL}) or Refs.\,\cite{Hiasa1, Kaur1}), and thus, it is natural to expect that the effect of $Q$ becomes larger as $\delta N$ increases.
In general, the magnitude of $Q$ is not commensurate with that of $Q_0$, and hence, the role of $Q$ can interfere with that of $Q_0$, which is thought to lead to the upturn. 

Furthermore, the emergence of the critical end point of the low-field FOST seems to be closely related to the upturn in Fig.\,\ref{fig:shape_compare_deltaN1}.
It is found in our calculation that, in the parameter ($\nu$, $\lambda$) space, where $\nu$ and $\lambda$ are defined in Fig.\,\ref{fig:lattice_parameters} as parameters characterizing the vortex lattice unit cell, phases I and II correspond to two neighboring valleys to each other.
In the previous $\delta N=0$ case, these two valleys move to the same direction in the parameter space as the field, and hence $Q_0$, increases, and consequently, they do not merge with each other. 
On the other hand, in the present case, the effect of the nonzero $Q_0$ is weakened by the presence of the finite $Q$ in particular at lower temperatures, and hence, the structure in phase II, which is more strongly compressed in the $x$ direction than that in the phase I, starts to return to a more stretched structure at a $\delta N$-dependent value of the applied field.
Therefore, the above-mentioned two valleys tend to merge with each other, resulting in the disappearance of the low-field FOST between the I and II phases and thus in the critical end point.
In fact, the change of the lattice structure in Fig.\,\ref{fig:phase_diagram_deltaN1} from (E) to (D) and then to (F) can be naturally understood as being due to the field-induced compression in the $x$ direction and stretch in the $z$ direction.

In the higher-field region where the anisotropic triangular lattice is destabilized, the resulting structure (C) has the rectangular symmetry.
Interestingly, this phase III with the rectangular symmetry is wide, and, with no transition, the nodelike region with extremely small $|\Delta|$ becomes wider as the field grows.
This increase of the spatial modulation of $|\Delta|$ in phase III is a consequence of the roles of the odd LLs due to the enhancement of the PPB effect in the higher-field region.
Furthermore, at the high-field end, we have the narrow phase IV with highly anisotropic and modulating structures. 

\subsection{Calculation of LDOS}

\begin{figure}
\centering
\includegraphics[scale=0.4]{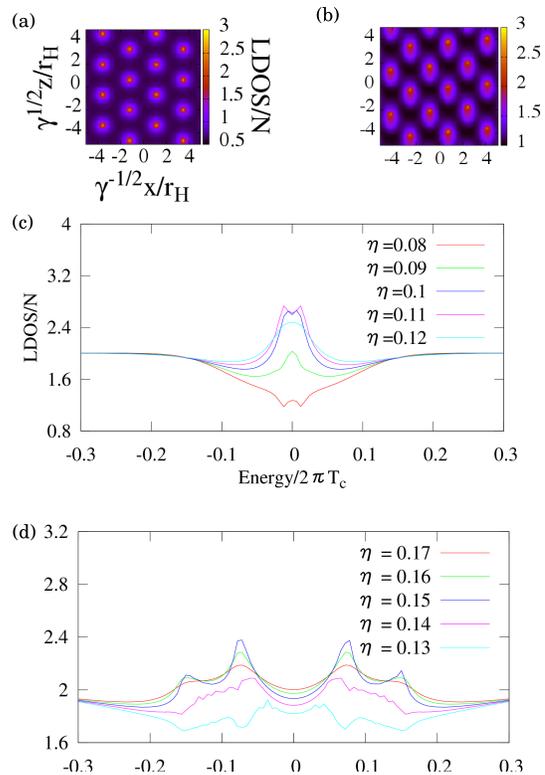}
\caption{(Color online) Panels (a) and (b) are the images of the local density of states (LDOS) at the excitation energy $E = 0$ of the states (F) and (D), respectively, in Fig.\,\ref{fig:phase_diagram_deltaN1}.
The smearing factor $\eta$ here is set at 0.1 and 0.15, respectively, in units of $2\pi T_c$.
Panels (c) and (d) are the graphs of the energy dependence of the LDOS at the vortex center of the same state as in (a) and (b), respectively, with different smearing factors $\eta$.
In (c), the spectrum is stable around $\eta/2\pi T_c = 0.1$ and has slightly split peaks.
In (d), the spectrum is stable around $\eta/2\pi T_c = 0.15$ and has widely split peaks.
}
\label{fig:LDOS_deltaN1}
\end{figure}

As available results for comparison with real experiments to be performed in future, the LDOS of vortex lattices have been examined, and their examples are shown in Fig.\,\ref{fig:LDOS_deltaN1}.
The smearing factor $\eta$ is fixed at the value where the $\eta$ dependence of the spectrum of the LDOS is moderate.

In the low-field region, there is a double peak structure with a narrow splitting around $E=0$ in the vortex core, and as the field increases, the splitting of the peaks grows wider.
This splitting seems to stem from the Zeeman effect \cite{Ichioka2}, because its width is nearly equal to the double of $\mu_sH$ ($= 0.4\cdot (H/H_{\mathrm{orb}}^{\mathrm{2D}})\cdot 2\pi T_c$ in the present cases).

Although the peaks are split due to the Zeeman effect, the spatial dependence of the LDOS at $E=0$ reflects that of $\left|\Delta\right|$ directly.
Thus, observation of these peaks around the vortex cores in STM experiments would lead to the verification of the compressing effect due to the finite $Q_0$ (i.e., the shift of the split Fermi surfaces induced by the interplay between the PPB effect and the ASOC).

We should comment, however, on the smearing factor $\eta$.
In the present approach, the value of $\eta/2\pi T_c$ is of $O(10^{-1})$ and is much larger than that in the methods used in such papers as Ref.\,\cite{Ichioka1} and Ref.\,\cite{Nagai1}, where $\eta/2\pi T_c = O(10^{-3})$.
Therefore, the results here are highly smeared, and the detailed information on the electronic structure may be lost.
Nevertheless, we believe that the essential structure is captured because there is a good correspondence between the vortex core and the peak of the LDOS.

We may be able to overcome this difficulty within the present framework, where the Eilenberger equation and the Landau level expansion are combined, by calculating the ``full solution,'' which is introduced in Ref.\,\cite{Adachi1}.
As is mentioned in Sec.\,\ref{sec:theory}, the validity of the approximate solution is not ensured in computing quantities related to the fine spatial structure of the system, which might result in the large $\eta$.
We can get over this point by examining the full solution, where the Fourier transform of the normal quasiclassical Green's function is employed instead of the approximation analogous to that used by Pesch \cite{Pesch}.
Development of the method to calculate the LDOS in this direction may be done in future works.

\section{Summary}
In this paper, possible vortex lattices in Rashba noncentrosymmetric superconductors under magnetic fields parallel to the basal plane have been studied based on the quasiclassical approach, and the obtained results have been compared with those in the previous GL approach \cite{Matsu,Hiasa1} neglecting the nonlocality \cite{AI03} in the quartic term of the GL free energy.

We have found that the overall field dependence of the vortex lattice structure in the realistic $\delta N \neq 0$ case remains unchanged even in the quasiclassical approach: 
The lattice structure is hexagonal in lower fields, while it is rectangular in higher fields.

However, we have also noticed that the details of the field dependence of the lattice structure are significantly changed.
First of all, the lattice structure at the high-field end is significantly changed compared with the corresponding GL result because of our proper treatment of the nonlocality of the quartic term in the free energy with respect to the superconducting order parameter.
The need for such treatment comes from the strong PPB effect.
Due to this effect, higher LLs, the effect of which is known to make the lattice structure complicated in the context of centrosymmetric superconductors \cite{Klein1}, play a more important role in the \textit{higher}-field region.
Meanwhile, their spatial variation is more intense than that of the lowest LL.
Therefore, the nonlocality has to be dealt with appropriately in such region.
Second, we have found a critical end point of a first-order structural transition line at a low temperature and an intermediate field, where the previous GL approach \cite{Matsu,Hiasa1} did not give any reliable result.
Furthermore, its appearance has been argued to be a reflection of the helical phase modulation \cite{Kaur1} which can be directly seen only in the vortex free limit.

Moreover, in the present work, we have been able to clarify that the origin of the complex field-dependent structural changes of the vortex lattice consists in the anisotropic compression of the lattice occurring as a consequence of the relative shift of the two Fermi surfaces due to the interplay between the PPB and the lack of the inversion symmetry.
In addition, we have also examined the LDOS in such vortex lattice structures.
We hope that, through some STM experiments, the strange field dependencies of the vortex lattice structure would be verified by measuring the LDOS in Rashba superconductors.

\begin{acknowledgements}
One of the authors (R.I.) was financially supported by a Grant-in-Aid for Scientific Research (No.\,25400368) from MEXT, Japan. 
\end{acknowledgements}

\appendix

\section{Diagonalization of $\mathcal{H}_{\mathrm{single}}$}
For convenience, we define here again the noninteracting part of the Hamiltonian $\mathcal{H}_{\mathrm{single}}$ and its concomitant quantities:
\begin{gather}
\mathcal{H}_{\mathrm{single}} = \sum_{\bm{k},\alpha,\beta}c^{\dag}_{\bm{k}\alpha}[\varepsilon_{\bm{k}}\sigma_0 + \zeta \bm{g}_{\bm{k}}\cdot\bm{\sigma}]_{\alpha,\beta}c_{\bm{k}\beta} \notag\\
+ \int d^3\bm{r}\sum_{\alpha\beta}c^{\dag}_{\alpha}(\bm{r})\mu_s\bm{B}(\bm{r})\cdot\bm{\sigma}_{\alpha,\beta}c_{\beta}(\bm{r}),
\end{gather}
where $c_{\bm{k}\alpha}$ is the annihilation operator of an electron with momentum $\bm{k}$ and spin $\alpha (=\uparrow,\downarrow)$, $c_{\alpha}(\bm{r})$ is the counterpart at the position $\bm{r}$ in the real space, and the $\sigma_{\mu}$'s ($\mu=0,1,2,3$) are the Pauli matrices. 
As to the centrosymmetric part of the quasiparticle dispersion $\varepsilon_{\bm{k}}$, the quasi two-dimensional form is assumed:
\begin{equation}
\varepsilon_{\bm{k}} = \frac{1}{2m}(k_x^2 + k_y^2) + J(1-\cos{k_zd}),
\end{equation}
where $m$ is the effective mass of a quasiparticle, and $d$ is the lattice constant in the $c$-axis direction; 
\begin{equation}
\bm{g}(\bm{k}) = \frac{\bm{k}_\perp \times\hat{\bm{z}}}{k_{\mathrm{F}}}, 
\end{equation}
is the ($g$) vector expressing the antisymmetric spin-orbit coupling (ASOC) of Rashba type, $\bm{k}_\perp=\bm{k} - k_z \hat{\bm{z}}$ is the two-dimensional wave vector, $k_{\mathrm{F}} = \sqrt{2mE_{\mathrm{F}}}$, $E_{\rm F}$ is the bare Fermi energy, $\hat{\bm{z}}$ is the unit vector in the direction of the broken inversion symmetry, and $\zeta$ is the strength of the ASOC. 
Throughout this paper, the $xy$ plane is the basal plane for the broken inversion symmetry, and $J$ is the interplane coupling constant.
In addition, $\mu_s$ is the magnetic moment of the spin of a quasiparticle, and $\bm{B}$ is the magnetic flux density. 

In the absence of magnetic fields, $\mathcal{H}_{\mathrm{single}}$ can be diagonalized by using the matrix 
\begin{equation}
U(\bm{k}) = \frac{\sigma_0 + i(\cos{\phi_{\bm{k}}}\sigma_1 + \sin{\phi_{\bm{k}}}\sigma_2)}{\sqrt{2}}
\end{equation}
($\phi_{\bm{k}} = \mathrm{tan}^{-1}\frac{k_y}{k_x}$), namely, by introducing the new field operator
\begin{equation}
\tilde{c}_{a\bm{k}} = U_{a\alpha}(\bm{k})c_{\alpha\bm{k}}. \\
\end{equation}
Then, $\mathcal{H}_{\mathrm{single}}$ with $\bm{B} = \bm{0}$ becomes
\begin{equation}
\mathcal{H}_{\mathrm{single}} = \sum_{\bm{k}}\{\varepsilon_{1\bm{k}}\tilde{c}^{\dag}_{1\bm{k}}\tilde{c}_{1\bm{k}} + \varepsilon_{2\bm{k}}\tilde{c}^{\dag}_{2\bm{k}}\tilde{c}_{2\bm{k}}\},
\end{equation}
where
\begin{equation}
\varepsilon_{a\bm{k}} = \varepsilon_{\bm{k}} + (-1)^{a+1}\zeta |\bm{g}_{\bm{k}}|.
\end{equation}

In the presence of a homogeneous magnetic field $\bm{H}$, the existence of the Zeeman energy term does not permit the precise diagonalization by using $U(\bm{k})$.
Nevertheless, if the temperature $T$ and the magnitude of the Zeeman energy $\mu_sH$ are sufficiently small compared to the strength of the ASOC $\zeta$, the interband mixing caused by the Zeeman term can be quantitatively neglected, so that $\mathcal{H}_{\mathrm{single}}$ simply becomes 
\begin{align}
\mathcal{H}_{\mathrm{single}} &= \sum_{\bm{k}}\left\{(\varepsilon_{1\bm{k}} + \mu_s\hat{\bm{g}}_{\bm{k}}\cdot\bm{H})\tilde{c}^{\dag}_{1\bm{k}}\tilde{c}_{1\bm{k}}\right.\notag\\
&\left.+ (\varepsilon_{2\bm{k}} - \mu_s\hat{\bm{g}}_{\bm{k}}\cdot\bm{H})\tilde{c}^{\dag}_{2\bm{k}}\tilde{c}_{2\bm{k}}\right\},
\end{align}
in which the momentum $\bm{Q}_0$ giving the shift between the two Fermi surfaces is given by 
\begin{equation}
\bm{Q}_0 = \frac{\mu_s H}{v_{\mathrm{F}}}\hat{\bm{x}}\label{eq:def_Q0_appendix}
\end{equation}
because
\begin{equation}
\mu_s\hat{\bm{g}}_{\bm{k}}\cdot\bm{H} \simeq -\bm{v}_{\mathrm{F}}\cdot\frac{\mu_s H}{v_{\mathrm{F}}}\hat{\bm{x}}
\end{equation}
is satisfied near the Fermi surface, where $\bm{v}_{\mathrm{F}}$ is the Fermi velocity.

\section{Derivation of the Eilenberger Equation}

Throughout the present paper, we use the mean field approximation 
\begin{equation}
\mathcal{H} \simeq \mathcal{H}_{\mathrm{single}} - \frac{V}{2}\sum_{\bm{q}}\left( \Delta_{\bm{q}}^{\ast}\Psi_{\bm{q}} + \Psi^{\dag}_{\bm{q}}\Delta_{\bm{q}}\right) + \frac{V}{g}\sum_{\bm{q}}|\Delta_{\bm{q}}|^2
\end{equation}
for the Hamiltonian ${\cal H}$, where $g(>0)$ is the coupling constant, $V$ is the volume of the system, and
\begin{equation}
\Psi_{\bm{q}}=\frac{1}{V}\sum_{\bm{k},\alpha,\beta}c_{-\bm{k}+\bm{q}/2,\alpha}(-i\sigma_2)_{\alpha\beta}c_{\bm{k}+\bm{q}/2,\beta}
\end{equation}
is the field operator of a spin-singlet $s$-wave Cooper pair with the total momentum $\bm{q}$. The component $\Delta_{\bm{q}}$ with the momentum $\bm{q}$ of the order parameter $\Delta(\bm{r})$ is given by 
\begin{equation}
\Delta_{\bm{q}} = -\frac{g}{2}\left<\Psi_{\bm{q}}\right>_{eq}, \label{eq:def_order_parameter}
\end{equation}
where 
$\Delta(\bm{r}) = \sum_{\bm{q}}e^{i\bm{q}\cdot\bm{r}}\Delta_{\bm{q}}$.
Here, $\left<\hat{X}\right>_{eq}$ is the grand canonical ensemble average of an arbitrary operator $\hat{X}$ under the Hamiltonian ${\cal H}$.

As usual, the Gor'kov Green's functions are defined as
\begin{eqnarray}
G_{\alpha\beta}(\bm{r}_1,\bm{r}_2;\tau_1-\tau_2) = -\left< T_{\tau} c_{\alpha}(\bm{r}_1,\tau_2) \bar{c}_{\beta}(\bm{r}_2,\tau_2) \right>_{eq},\notag\\
\bar{G}_{\alpha\beta}(\bm{r}_1,\bm{r}_2;\tau_1-\tau_2) = -\left< T_{\tau} \bar{c}_{\alpha}(\bm{r}_1,\tau_1) c_{\beta}(\bm{r}_2,\tau_2) \right>_{eq},\notag\\
F_{\alpha\beta}(\bm{r}_1,\bm{r}_2;\tau_1-\tau_2) = -\left< T_{\tau} c_{\alpha}(\bm{r}_1,\tau_1) c_{\beta}(\bm{r}_2,\tau_2) \right>_{eq},\notag\\
\bar{F}_{\alpha\beta}(\bm{r}_1,\bm{r}_2;\tau_1-\tau_2) = -\left< T_{\tau} \bar{c}_{\alpha}(\bm{r}_1,\tau_1) \bar{c}_{\beta}(\bm{r}_2,\tau_2) \right>_{eq},\notag\\
\end{eqnarray}
where $T_{\tau}$ denotes the imaginary time ordering operation, and 
\begin{eqnarray}
c_{\alpha}(\tau) =& e^{(\mathcal{H}-\mu\mathcal{N})\tau}c_{\alpha}e^{-(\mathcal{H}-\mu\mathcal{N})\tau},\notag\\
\bar{c}_{\alpha}(\tau) =& e^{(\mathcal{H}-\mu\mathcal{N})\tau}c_{\alpha}^{\dag}e^{-(\mathcal{H}-\mu\mathcal{N})\tau}.
\end{eqnarray}
Here, $\mu$ is the chemical potential, and $\mathcal{N} = \sum_{\bm{k},\alpha}c^{\dag}_{\bm{k}\alpha}c_{\bm{k}\alpha}$ is the particle number.
By taking derivatives of the Gor'kov Green's functions with respect to the imaginary time, the following left- and right-sided Gor'kov equations are obtained:
\begin{widetext}
\begin{align}
-\partial_{\tau_1} G_{\alpha\beta}(\bm{r}_1,\bm{r}_2;\tau_1-\tau_2)=&\delta^3(\bm{r}_1-\bm{r}_2)\delta(\tau_1-\tau_2)\delta_{\alpha\beta}\notag\\
&\!\!\!\!\!\!\!\!\!\!\!\!\!\!\!\!\!\!\!\!\!\!\!\!\!\!\!\!\!\!+\sum_{\gamma}[\xi(-i\nabla_1+e\bm{A}(\bm{r}_1))+\mu_s\bm{\sigma}\cdot\bm{B}(\bm{r}_1)]_{\alpha\gamma}G_{\gamma\beta}(\bm{r}_1,\bm{r}_2;\tau_1-\tau_2)-\sum_{\gamma}\Delta_{\alpha\gamma}(\bm{r}_1)\bar{F}_{\gamma\beta}(\bm{r}_1,\bm{r}_2;\tau_1-\tau_2),\notag
\end{align}
\vspace{-4ex}
\begin{align}
-\partial_{\tau_1} \bar{G}_{\alpha\beta}(\bm{r}_1,\bm{r}_2;\tau_1-\tau_2)=&\delta^3(\bm{r}_1-\bm{r}_2)\delta(\tau_1-\tau_2)\delta_{\alpha\beta}\notag\\
&\!\!\!\!\!\!\!\!\!\!\!\!\!\!\!\!\!\!\!\!\!\!\!\!\!\!\!\!\!\!-\sum_{\gamma}[\xi(i\nabla_1+e\bm{A}(\bm{r}_1))+\mu_s\bm{\sigma}\cdot\bm{B}(\bm{r}_1)]^{\mathrm{T}}_{\alpha\gamma}\bar{G}_{\gamma\beta}(\bm{r}_1,\bm{r}_2;\tau_1-\tau_2)-\sum_{\gamma}\Delta^{\dag}_{\alpha\gamma}(\bm{r}_1)F_{\gamma\beta}(\bm{r}_1,\bm{r}_2;\tau_1-\tau_2),\notag
\end{align}
\vspace{-4ex}
\begin{align}
-\partial_{\tau_1} F_{\alpha\beta}(\bm{r}_1,\bm{r}_2;\tau_1-\tau_2)=&\notag\\
&\!\!\!\!\!\!\!\!\!\!\!\!\!\!\!\!\!\!\!\!\!\!\!\!\!\!\!\!\!\!\sum_{\gamma}[\xi(-i\nabla_1+e\bm{A}(\bm{r}_1))+\mu_s\bm{\sigma}\cdot\bm{B}(\bm{r}_1)]_{\alpha\gamma}F_{\gamma\beta}(\bm{r}_1,\bm{r}_2;\tau_1-\tau_2)-\sum_{\gamma}\Delta_{\alpha\gamma}(\bm{r}_1)\bar{G}_{\gamma\beta}(\bm{r}_1,\bm{r}_2;\tau_1-\tau_2),\notag
\end{align}
\vspace{-4ex}
\begin{align}
-\partial_{\tau_1} \bar{F}_{\alpha\beta}(\bm{r}_1,\bm{r}_2;\tau_1-\tau_2)=&\notag\\
&\!\!\!\!\!\!\!\!\!\!\!\!\!\!\!\!\!\!\!\!\!\!\!\!\!\!\!\!\!\!-\sum_{\gamma}[\xi(i\nabla_1+e\bm{A}(\bm{r}_1))+\mu_s\bm{\sigma}\cdot\bm{B}(\bm{r}_1)]^{\mathrm{T}}_{\alpha\gamma}\bar{F}_{\gamma\beta}(\bm{r}_1,\bm{r}_2;\tau_1-\tau_2)-\sum_{\gamma}\Delta^{\dag}_{\alpha\gamma}(\bm{r}_1)G_{\gamma\beta}(\bm{r}_1,\bm{r}_2;\tau_1-\tau_2)
\label{GEQS_left}
\end{align}
and
\begin{align}
\partial_{\tau_2} G_{\alpha\beta}(\bm{r}_1,\bm{r}_2;\tau_1-\tau_2)=&\delta^3(\bm{r}_1-\bm{r}_2)\delta(\tau_1-\tau_2)\delta_{\alpha\beta}\notag\\
&\!\!\!\!\!\!\!\!\!\!\!\!\!\!\!\!\!\!\!\!\!\!\!\!\!\!\!\!\!\!+\sum_{\gamma}G_{\alpha\gamma}(\bm{r}_1,\bm{r}_2;\tau_1-\tau_2)[\xi(-i\nabla_2+e\bm{A}(\bm{r}_2))+\mu_s\bm{\sigma}\cdot\bm{B}(\bm{r}_2)]_{\gamma\beta}-\sum_{\gamma}F_{\alpha\gamma}(\bm{r}_1,\bm{r}_2;\tau_1-\tau_2)\Delta^{\dag}_{\gamma\beta}(\bm{r}_2),\notag
\end{align}
\vspace{-4ex}
\begin{align}
\partial_{\tau_2} \bar{G}_{\alpha\beta}(\bm{r}_1,\bm{r}_2;\tau_1-\tau_2)=&\delta^3(\bm{r}_1-\bm{r}_2)\delta(\tau_1-\tau_2)\delta_{\alpha\beta}\notag\\
&\!\!\!\!\!\!\!\!\!\!\!\!\!\!\!\!\!\!\!\!\!\!\!\!\!\!\!\!\!\!-\sum_{\gamma}\bar{G}_{\alpha\gamma}(\bm{r}_1,\bm{r}_2;\tau_1-\tau_2)[\xi(i\nabla_2+e\bm{A}(\bm{r}_2))+\mu_s\bm{\sigma}\cdot\bm{B}(\bm{r}_2)]^{\mathrm{T}}_{\gamma\beta}-\sum_{\gamma}\bar{F}_{\alpha\gamma}(\bm{r}_1,\bm{r}_2;\tau_1-\tau_2)\Delta_{\gamma\beta}(\bm{r}_2),\notag
\end{align}
\vspace{-4ex}
\begin{align}
\partial_{\tau_2} F_{\alpha\beta}(\bm{r}_1,\bm{r}_2;\tau_1-\tau_2)=&\notag\\
&\!\!\!\!\!\!\!\!\!\!\!\!\!\!\!\!\!\!\!\!\!\!\!\!\!\!\!\!\!\!-\sum_{\gamma}F_{\alpha\gamma}(\bm{r}_1,\bm{r}_2;\tau_1-\tau_2)[\xi(i\nabla_2+e\bm{A}(\bm{r}_2))+\mu_s\bm{\sigma}\cdot\bm{B}(\bm{r}_2)]^{\mathrm{T}}_{\gamma\beta}-\sum_{\gamma}G_{\alpha\gamma}(\bm{r}_1,\bm{r}_2;\tau_1-\tau_2)\Delta_{\gamma\beta}(\bm{r}_2),\notag
\end{align}
\vspace{-4ex}
\begin{align}
\partial_{\tau_2} \bar{F}_{\alpha\beta}(\bm{r}_1,\bm{r}_2;\tau_1-\tau_2)=&\notag\\
&\!\!\!\!\!\!\!\!\!\!\!\!\!\!\!\!\!\!\!\!\!\!\!\!\!\!\!\!\!\!\sum_{\gamma}\bar{F}_{\alpha\gamma}(\bm{r}_1,\bm{r}_2;\tau_1-\tau_2)[\xi(-i\nabla_2+e\bm{A}(\bm{r}_2))+\mu_s\bm{\sigma}\cdot\bm{B}(\bm{r}_2)]_{\gamma\beta}-\sum_{\gamma}\bar{G}_{\alpha\gamma}(\bm{r}_1,\bm{r}_2;\tau_1-\tau_2)\Delta^{\dag}_{\gamma\beta}(\bm{r}_2),
\label{GEQS_right}
\end{align}
\end{widetext}
where $-e$ is the electronic charge, $\bm{A}$ is the vector potential associated with $\bm{B}$,
\begin{equation}
\Delta_{\alpha\beta}(\bm{r}) = (-i\sigma_2)_{\alpha\beta}\Delta(\bm{r})
\end{equation}
and
\begin{equation}
\xi(\bm{k}) = (\varepsilon_{\bm{k}}- \mu)\sigma_0 + \zeta\bm{g}_{\bm{k}}\cdot\bm{\sigma}.
\end{equation}
Hereafter, we define the operation of $\nabla$ to an arbitrary function $h(\bm{r})$ from the right side as
\begin{equation}
h(\bm{r})\nabla = -\nabla h(\bm{r}).
\end{equation}

The Wigner representation of the Green's functions is 
\begin{align}
X_{\alpha\beta}&(\bm{r};\bm{k},\omega_n) =\notag\\
&\!\!\!\!\!\!\!\!\!\!\int d^3\bm{r}^{\prime} e^{-i\bm{k}\cdot\bm{r}^{\prime}} \int_0^{1/T}d\tau e^{i\omega_n\tau} X_{\alpha\beta}(\bm{r}+\bm{r}^{\prime}/2, \bm{r}-\bm{r}^{\prime}/2; \tau)
\end{align}
for $X = G,\bar{G},F,\bar{F}$. 
By Fourier-transforming Eqs.\,(\ref{GEQS_left}) and (\ref{GEQS_right}) and neglecting the higher-order terms with respect to $1/k_{\mathrm{F}}\xi_x$ and $1/k_{\mathrm{F}}\xi_z$ ($\xi_x$ and $\xi_z$ are the coherent lengths in the $x$ and $z$ directions), the left- and right-sided Gor'kov equations in the Wigner representation become
\begin{equation}
\check{G}^{-1}\check{G} = \check{G}\check{G}^{-1} = \check{1}
\label{LRGEQS}
\end{equation}
in the matrix form, where
\begin{align}
&\check{G}(\bm{r};\bm{k},\omega_n)=
\begin{pmatrix}
G(\bm{r};\bm{k},\omega_n) & F(\bm{r};\bm{k},\omega_n)\\
-\bar{F}(\bm{r};\bm{k},\omega_n) & -\bar{G}(\bm{r};\bm{k},\omega_n)
\end{pmatrix},
\end{align}
\begin{widetext}
\begin{align}
\check{G}^{-1}&(\bm{r};\bm{k},\omega_n)= \notag\\
&
\begin{pmatrix}
i\omega_n\sigma_0-[\xi(\bm{k})+\bm{v}(\bm{k})\cdot\bm{\Pi}/2+\mu_s\bm{\sigma}\cdot\bm{B}(\bm{r})] & i\sigma_2\Delta(\bm{r})\\
-i\sigma_2\Delta^{\ast}(\bm{r}) & -i\omega_n\sigma_0-[\xi^{\mathrm{T}}(-\bm{k})+\bm{v}^{\mathrm{T}}(-\bm{k})\cdot\bm{\Pi}^{\ast}/2+\mu_s\bm{\sigma}^{\mathrm{T}}\cdot\bm{B}(\bm{r})]
\end{pmatrix}
\end{align}
\end{widetext}
and $\check{1}$ is the $4\times 4$ identity matrix.
Here,
\begin{eqnarray}
\bm{\Pi} =& -i\nabla +2e\bm{A},\notag\\
\bm{\Pi}^{\ast} =& i\nabla + 2e\bm{A},
\end{eqnarray}
\begin{align}
X=
\begin{pmatrix}
X_{\uparrow\uparrow} & X_{\uparrow\downarrow}\\
X_{\downarrow\uparrow} & X_{\downarrow\downarrow}
\end{pmatrix}
\ (X = G,\bar{G},F,\bar{F})
\end{align}
and
\begin{equation}
\bm{v}(\bm{k}) = \nabla_{\bm{k}}\xi(\bm{k}).
\end{equation}
The subtraction of the two equations in Eq.\,(\ref{LRGEQS}) leads to the following equation 
\begin{equation}
[\check{G}^{-1},\check{G}]=0.\label{eq:left-right-gorkov}
\end{equation}

To obtain Gor'kov equations in a more useful form, the transformation with the matrix
\begin{equation}
\check{U}(\bm{k}) =
\begin{pmatrix}
U(\bm{k}) & 0 \\
0 & U(-\bm{k})^{\ast}
\end{pmatrix}
\end{equation}
is considered, namely, the Green's functions and its inverse operator in the transformed new representation are defined as
\begin{equation}
\check{G}^{\prime} = \check{U}(\bm{k})\check{G}\check{U}^{\dag}(\bm{k}),
\label{eq:green_func_trans}
\end{equation}
and
\begin{equation}
\check{G}^{\prime -1} = \check{U}(\bm{k})\check{G}^{-1}\check{U}^{\dag}(\bm{k}),
\end{equation}
and further, we put
\begin{equation}
\check{G}^{\prime} =
\begin{pmatrix}
G_{1} & G_{12} & F_{1} & F_{12} \\
G_{21} & G_{2} & F_{21} & F_{2} \\
-\bar{F}_{1} & -\bar{F}_{12} & -\bar{G}_{1} & -\bar{G}_{12} \\
-\bar{F}_{21} & -\bar{F}_{2} & -\bar{G}_{21} & -\bar{G}_{2} \\
\end{pmatrix}.
\end{equation}
Suppose here that the length scales on any inhomogeneity are sufficiently longer than $k_{\mathrm{F}}^{-1}$.
Then, $X_{1}$ and $X_{2}$ are interpreted as the intraband Green's functions of the bands 1 and 2 at $\bm{r}$, while $X_{12}$ and $X_{21}$ are interpreted as the interband ones ($X=G,\bar{G},F,\bar{F}$).
Here, we neglect the off-diagonal elements by assuming that $T_c$, the critical temperature at zero field, and $\mu_sH$, the magnitude of the Zeeman energy, are much smaller than $\zeta$, the strength of the ASOC (see the main text). 
Then, Eq.\,(\ref{eq:left-right-gorkov}) becomes
\begin{equation}
[\check{G}_a^{-1},\check{G}_a]=0,\label{eq:left-right-gorkov-band-rep}
\end{equation}
where
\begin{equation}
\check{G}_a=
\begin{pmatrix}
G_a & F_a\\
-\bar{F}_a & -\bar{G}_a
\end{pmatrix}
\end{equation}
and
\begin{widetext}
\begin{align}
\check{G}_a^{-1}&(\bm{r};\bm{k},\omega_n) = \notag\\
&
\begin{pmatrix}
i\omega_n-[\xi_a+\bm{v}_a\cdot\bm{\Pi}/2+(-1)^{a+1}\mu_s\hat{\bm{g}}_{\bm{k}}\cdot\bm{B}(\bm{r})] & w_{a}\Delta(\bm{r})\\
-w_{a}^{\ast}\Delta^{\ast}(\bm{r}) & -i\omega_n-[\xi_a-\bm{v}_a\cdot\bm{\Pi}^{\ast}/2-(-1)^{a+1}\mu_s\hat{\bm{g}}_{\bm{k}}\cdot\bm{B}(\bm{r})]
\end{pmatrix}.
\end{align}
\end{widetext}
Here,
\begin{equation}
\xi_a = \varepsilon_{a\bm{k}} - \mu,
\end{equation}
\begin{align}
w_a =& (-1)^aie^{i (-1)^a \phi_{\bm{k}}}
\end{align}
and
\begin{equation}
\begin{pmatrix}
\bm{v}_{1} & \bm{v}_{12} \\
\bm{v}_{21} & \bm{v}_{2}
\end{pmatrix}
=
U(\bm{k})
\begin{pmatrix}
\bm{v}_{\uparrow\uparrow}(\bm{k}) & \bm{v}_{\uparrow\downarrow}(\bm{k}) \\
\bm{v}_{\downarrow\uparrow}(\bm{k}) & \bm{v}_{\downarrow\downarrow}(\bm{k})
\end{pmatrix}
U^{\dag}(\bm{k}).
\end{equation}

\begin{figure}
\centering
\includegraphics{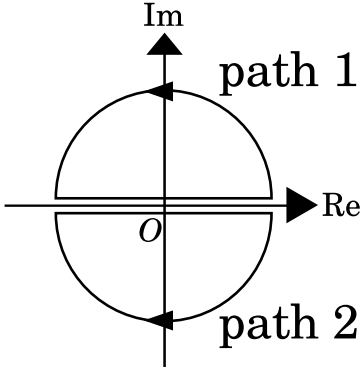}
\caption{Two paths of the complex integration $\oint$.}
\label{fig:path}
\end{figure}

In this representation, we define the {\it quasiclassical} Green's functions on each FS as follows:
\begin{align}
g_a(\bm{r};\hat{\bm{k}},\omega_n) = \oint \frac{d\xi_a}{\pi i} G_a(\bm{r};\bm{k},\omega_n),\\
\bar{g}_a(\bm{r};\hat{\bm{k}},\omega_n) = \oint \frac{d\xi_a}{\pi i} \bar{G}_a(\bm{r};\bm{k},\omega_n),\\
f_a(\bm{r};\hat{\bm{k}},\omega_n) = \oint \frac{d\xi_a}{\pi i} F_a(\bm{r};\bm{k},\omega_n),\\
\bar{f}_a(\bm{r};\hat{\bm{k}},\omega_n) = \oint \frac{d\xi_a}{\pi i} \bar{F}_a(\bm{r};\bm{k},\omega_n),
\end{align}
and their matrix form as
\begin{equation}
\check{g}_a=
\begin{pmatrix}
g_a & f_a\\
-\bar{f}_a & -\bar{g}_a
\end{pmatrix}.
\end{equation}
Here, the complex integration $\oint$ represents the average of the two contour integrals along the paths 1 and 2 illustrated in Fig.\,\ref{fig:path}.
The integration of Eq.\,(\ref{eq:left-right-gorkov-band-rep}) with respect to $\xi_a$ leads to the Eilenberger equation: 
\begin{equation}
\left[\left.\check{G}^{-1}_a\right|_{\bm{k}=\bm{k}_{\mathrm{F}a}},\check{g}_a\right]=0,\label{eq:eilenberger_matrix}
\end{equation}
where $\bm{k}_{\mathrm{F}a}$ is the Fermi wave vector of the $a$th band, and we also use the fact that every Green's function has a sharp peak with the width $|\Delta|$ around $\xi_a=0$.

\section{Conditions on $\check{g}_a$}
Because of the subtraction which leads to Eq.\,(\ref{eq:left-right-gorkov}), some information is lost.
Actually, we cannot determine the quasiclassical Green's functions uniquely based only on Eq.\,(\ref{eq:eilenberger_matrix}).
This information, however, can be recovered with the following two conditions:
\begin{align}
\check{g}_a^2 = \check{1},\,\,\,\,\,\,  \mathrm{sgn}\,\mathrm{Re}\,g_a = -\mathrm{sgn}\,\omega_n,
\end{align}
where $\check{1}$ is the $2\times 2$ identity matrix.
In this section, we derive these conditions based on some assumptions.

According to the Eilenberger equation (\ref{eq:eilenberger_matrix}), we have the relations 
\begin{gather}
\bm{v}_{\mathrm{F}a}\cdot\nabla(g_a - \bar{g}_a) = 0,\label{eq:eilenberger_invariant_1}\\
\bm{v}_{\mathrm{F}a}\cdot\nabla\left\{\left(\frac{g_a + \bar{g}_a}{2}\right)^2 - f_a\bar{f}_a\right\} = 0,\label{eq:eilenberger_invariant_2}
\end{gather}
where $\bm{v}_{\mathrm{F}a} = \left.\bm{v}_a\right|_{\bm{k}=\bm{k}_{\mathrm{F}a}}$.

In the spatially homogeneous case, the solutions of the Gor'kov equations in Eqs.\,(\ref{GEQS_left}) or in Eqs.\,(\ref{GEQS_right}) lead to the quasiclassical Green's functions
\begin{gather}
g_a = \bar{g}_a = -\frac{\omega_n}{\sqrt{\omega_n^2 + |w_a\Delta|^2}}\label{eq:homogenious_g},\\
f_a = -\frac{w_a\Delta}{i\sqrt{\omega_n^2 + |w_a\Delta|^2}}, \bar{f}_a = -\frac{(w_a\Delta)^{\ast}}{i\sqrt{\omega_n^2 + |w_a\Delta|^2}},\label{eq:homogenious_f}
\end{gather}
which obey 
\begin{equation}
\check{g}_a^2 = \check{1}. 
\label{eq:normalization}
\end{equation}
These are the physical solution of the Eilenberger equation (\ref{eq:eilenberger_matrix}).

In the cases with spatial inhomogeneity, it is assumed that the system can be smoothly transformed to the homogeneous state towards infinity so that the normalization condition (\ref{eq:normalization}) is still valid according to Eqs.\,(\ref{eq:eilenberger_invariant_1}) and (\ref{eq:eilenberger_invariant_2}).
Therefore, generally, 
\begin{equation}
g_a = \bar{g}_a = -\sqrt{1 + f_a\bar{f}_a}
\label{rootg}
\end{equation}
is also valid. Strictly speaking, the branch of the root in Eq.\,(\ref{rootg}) cannot be determined based only on Eq.\,(\ref{eq:normalization}). However, we note that the inequality
\begin{equation}
|f_a\bar{f}_a| < 1
\label{eq:f_times_fbar}
\end{equation}
is satisfied in the homogeneous case, and that the averaged magnitude of the superconducting energy gap $|\Delta|$ becomes smaller due to a spatial inhomogeneity of $\Delta$.
Thus, the inequality (\ref{eq:f_times_fbar}) should remain valid in vortex states so that the same branch as in the homogeneous case may be chosen.

\section{Symmetry Relations of Quasiclassical Green's Functions}
In this section, we derive some symmetry relations connecting one quasiclassical Green's function with another.

In the Wigner representation, the Gor'kov Green's functions have the symmetry relations
\begin{eqnarray}
F_{\alpha\beta}(\bm{r};\bm{k},\omega_n) &=& -F_{\beta\alpha}(\bm{r};-\bm{k},-\omega_n),\notag\\
\bar{F}_{\alpha\beta}(\bm{r};\bm{k},\omega_n) &=& -\bar{F}_{\beta\alpha}(\bm{r};-\bm{k},-\omega_n),\notag\\
F_{\alpha\beta}(\bm{r};\bm{k},\omega_n) &=& \bar{F}_{\beta\alpha}(\bm{r};\bm{k},-\omega_n)^{\ast},\notag\\
G_{\alpha\beta}(\bm{r};\bm{k},\omega_n) &=& -\bar{G}_{\beta\alpha}(\bm{r};-\bm{k},-\omega_n),\notag\\
G_{\alpha\beta}(\bm{r};\bm{k},\omega_n) &=& G_{\beta\alpha}(\bm{r};\bm{k},-\omega_n)^{\ast},\notag\\
\bar{G}_{\alpha\beta}(\bm{r};\bm{k},\omega_n) &=& \bar{G}_{\beta\alpha}(\bm{r};\bm{k},-\omega_n)^{\ast} 
\end{eqnarray}
following from their definition. In other words, with the use of the transformation (\ref{eq:green_func_trans}), we have 
\begin{eqnarray}
F_{a}(\bm{r};\bm{k},\omega_n) &=& -F_{a}(\bm{r};-\bm{k},-\omega_n),\notag\\
\bar{F}_{a}(\bm{r};\bm{k},\omega_n) &=& -\bar{F}_{a}(\bm{r};-\bm{k},-\omega_n),\notag\\
F_{a}(\bm{r};\bm{k},\omega_n) &=& \bar{F}_{a}(\bm{r};\bm{k},-\omega_n)^{\ast},\notag\\
G_{a}(\bm{r};\bm{k},\omega_n) &=& -\bar{G}_{a}(\bm{r};-\bm{k},-\omega_n),\notag\\
G_{a}(\bm{r};\bm{k},\omega_n) &=& G_{a}(\bm{r};\bm{k},-\omega_n)^{\ast},\notag\\
\bar{G}_{a}(\bm{r};\bm{k},\omega_n) &=& \bar{G}_{a}(\bm{r};\bm{k},-\omega_n)^{\ast}.
\end{eqnarray}
Integrating these equations with respect to $\xi_a$ and using the fact that $g_a = \bar{g}_a$ lead to 
\begin{eqnarray}
f_a(\bm{r};\bm{k},\omega_n) &=& -f_a(\bm{r};-\bm{k},-\omega_n) ,\notag\\
\bar{f}_a(\bm{r};\bm{k},\omega_n) &=& -\bar{f}_a(\bm{r};-\bm{k},-\omega_n) ,\notag\\
f_a(\bm{r};\bm{k},\omega_n) &=& -\bar{f}_a(\bm{r};\bm{k},-\omega_n)^{\ast} ,\notag\\
g_a(\bm{r};\bm{k},\omega_n) &=& -g_a(\bm{r};-\bm{k},-\omega_n) ,\notag\\
g_a(\bm{r};\bm{k},\omega_n) &=& -g_a(\bm{r};\bm{k},-\omega_n)^{\ast}, \label{eq:rashba_fundsym}
\end{eqnarray}
from which useful relations
\begin{eqnarray}
f_a(\bm{r};\bm{k},\omega_n) &=& \bar{f}_a(\bm{r};-\bm{k},\omega_n)^{\ast} ,\notag\\
g_a(\bm{r};\bm{k},\omega_n) &=& g_a(\bm{r};-\bm{k},\omega_n)^{\ast} \label{eq:frequently_used_sym}
\end{eqnarray}
are obtained. 

We often use these relations in this paper when summands in $\bm{k}$ or $\omega_n$ summations include the quasiclassical Green's functions.

\section{Approximation on Fermi Velocity}
\begin{figure}
\centering
\includegraphics[scale=0.6]{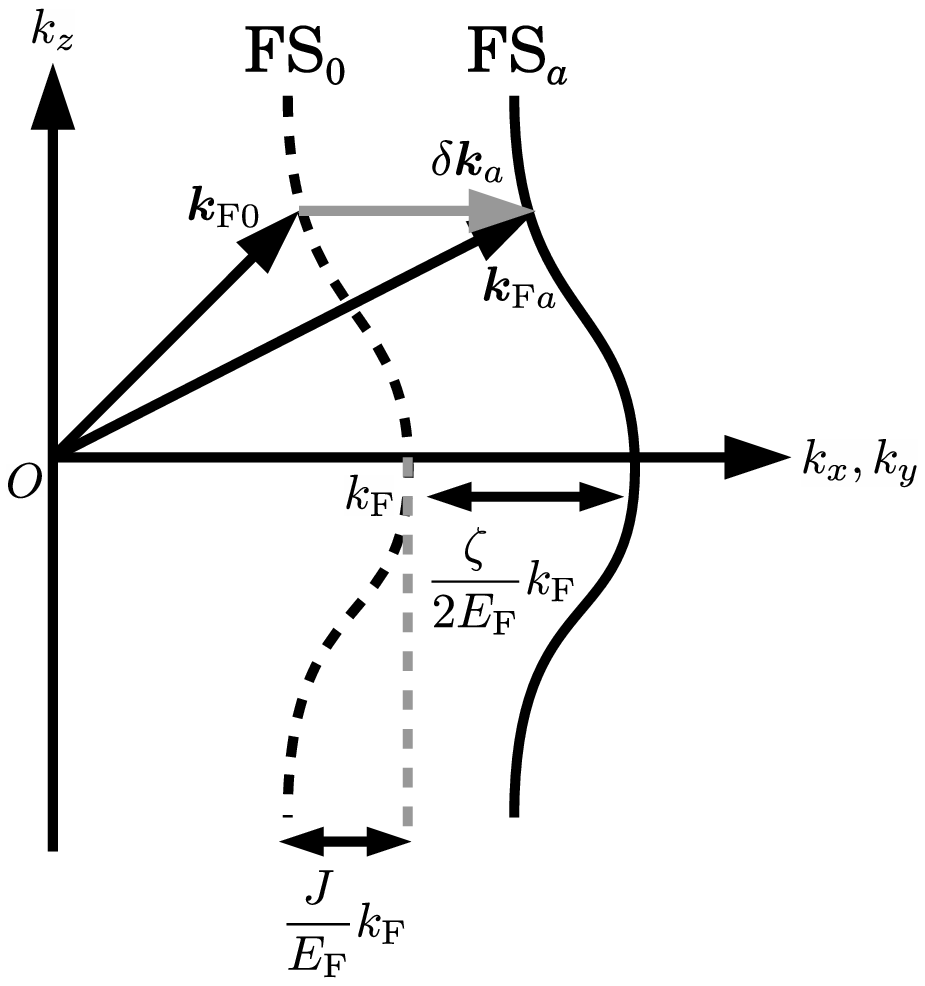}
\caption{Relations among the three vectors in Eq.\,(\ref{eq:def_delta_k}). $\mathrm{FS}_0$ denotes the Fermi surface of the bare band, while $\mathrm{FS}_a$'s ($a=1$, $2$) express the two Fermi surfaces split by the ASOC of Rashba type. In the figure, $\mathrm{FS}_a$ is represented by that of band 2.}
\label{fig:def_kF}
\end{figure}

The velocity in the band $a$ is given, following its definition, by 
\begin{equation}
\bm{v}_a = \bm{v}_0 + (-1)^{a+1}\frac{\zeta}{k_{\mathrm{F}}}\hat{\bm{k}}_{\perp},\label{eq:new_fermi_velocity}
\end{equation}
where
\begin{equation}
\bm{v}_0 = \nabla_{\bm{k}}\varepsilon_{\bm{k}}, 
\end{equation}
and
\begin{equation}
\hat{\bm{k}}_{\perp} = \bm{k}_{\perp}/\left|\bm{k}_{\perp}\right|.
\end{equation}
Here, we put
\begin{equation}
\bm{k}_{\mathrm{F}a} = \bm{k}_{\mathrm{F}0} + \delta\bm{k}_a,\label{eq:def_delta_k}
\end{equation}
where $\bm{k}_{\mathrm{F}a}$ ($a=1$, $2$) and $\bm{k}_{\mathrm{F}0}$ are the Fermi wave vectors on the Fermi surfaces split by the ASOC and that on the bare band, respectively (see Fig.\,\ref{fig:def_kF}). 
Keeping
\begin{equation}
\frac{\delta\bm{k}_a}{k_{\mathrm{F}}} = O\left(\frac{\zeta}{E_{\mathrm{F}}}\right)
\end{equation}
in mind, we get
\begin{equation}
\bm{v}_{\mathrm{F}}\cdot\delta\bm{k}_a = -(-1)^{a+1}\zeta
\end{equation}
from $\mathcal{H}_{\mathrm{single}}$, where $\bm{v}_{\mathrm{F}} = \left.\bm{v}_0\right|_{\bm{k}=\bm{k}_{\mathrm{F0}}}$.
Hereafter, the terms of $O((\zeta/E_{\mathrm{F}})^2)$, $O((J/E_{\mathrm{F}})^2)$ and $O(J\zeta/E_{\mathrm{F}}^2)$ are neglected.
In this approximation,
\begin{equation}
\delta\bm{k}_a = -(-1)^{a+1}m\frac{\zeta}{k_{\mathrm{F}}}\hat{\bm{k}}_{\perp}.
\end{equation}
Substituting this expression to Eq.\,(\ref{eq:def_delta_k}) and using Eq.\,(\ref{eq:new_fermi_velocity}) lead to
\begin{equation}
\bm{v}_{\mathrm{F}a} = \bm{v}_{\mathrm{F}}.
\end{equation}

\section{Derivation of Free Energy}
The free energy measure from that in the normal phase, i.e., 
\begin{equation}
F = -T\ln\mathrm{tr}e^{-\beta\mathcal{H}} + T\ln\mathrm{tr}e^{-\beta\mathcal{H}|_{\Delta=0}}
\end{equation}
($\beta = 1/T$), is used in the text to determine the vortex lattice structure at each field and temperature.
Since obtaining a tractable expression of the free energy directly from the above expression is not easy, it is first rewritten by following the procedure based on the variational principle \cite{Eilenberger1} used by Eilenberger. 

For this purpose, the gap equation and the expression of the electric current density are needed. 
According to the definition of the order parameter (\ref{eq:def_order_parameter}) and the relation (\ref{eq:frequently_used_sym}), the gap equation is
\begin{equation}
\frac{1}{g}\Delta + 2\pi T \sum_{0 < \omega_n < \omega_c,a}\frac{N_a}{2}\left< iw_a^{\ast}f_a \right>_{\mathrm{FS}} = 0,
\end{equation}
where $N_a$ is the normal DOS on each FS, 
\begin{equation}
\left<h(\bm{k})\right>_{\mathrm{FS}} = \int_{-\pi/d}^{\pi/d}\frac{dk_z}{2\pi/d}\int_{0}^{2\pi}\frac{d\phi_{\bm{k}}}{2\pi}h(\bm{k})
\end{equation}
is the average over each FS for an arbitrary function $h(\bm{k})$, and $\omega_c$ is the frequency cutoff introduced to prevent the divergence of the summation. To treat $\omega_c$ implicitly, 
we transform this equation to
\begin{equation}
N \ln{\frac{T}{T_c}} \Delta + 2\pi T \sum_{\omega_n > 0,a}\frac{N_a}{2}\left< iw_a^{\ast}f_a + \frac{|w_a|^2\Delta}{\omega_n} \right>_{\mathrm{FS}} = 0
\label{eq:gap_equation}
\end{equation}
by introducing the mean-field transition temperature $T_c$ at zero field through the well-known relation 
\begin{equation}
\frac{1}{g} - 2\pi T N\sum_{0 < \omega_n < \omega_c}\frac{1}{\omega_n} = N\ln{\frac{T}{T_c}}.
\end{equation}
Here, $N$ is the average of the normal DOS on the two FSs.
The current density is obtained from the relation
\begin{equation}
\bm{j} = -\left<\frac{\delta\mathcal{H}}{\delta\bm{A}}\right>_{eq} = \bm{j}_{S} + \bm{j}_{M},\label{eq:current}
\end{equation}
where
\begin{equation}
\bm{j}_S = -e\sum_{\alpha,\beta}\left<c^{\dag}_{\alpha}(\bm{r})\bm{v}_{\alpha\beta}(-i\nabla + e\bm{A})c_{\beta}(\bm{r})\right>_{eq},
\end{equation}
\begin{equation}
\bm{j}_M = -\mu_s\nabla\times\sum_{\alpha,\beta}\left<c^{\dag}_{\alpha}(\bm{r})\bm{\sigma}_{\alpha\beta}c_{\beta}(\bm{r})\right>_{eq}. 
\end{equation}
The relative current density components measured from their normal counterparts are expressed in terms of the quasiclassical Green's functions as follows: 
\begin{align}
\varDelta\bm{j}_S =& \bm{j}_S - \left.\bm{j}_S\right|_{\Delta=0}\notag\\
=& -\frac{T}{V}\sum_{\bm{k},\omega_n,\alpha,\beta}e^{i\omega_n 0^{+}}e\bm{v}_{\alpha\beta}(\bm{k})\varDelta G_{\beta\alpha}(\bm{r};\bm{k},\omega_n) \notag\\
=& -ie2\pi T\sum_{\omega_n>0,a}N_a\left< \bm{v}_{a}(g_a + 1) \right>_{\mathrm{FS}},
\end{align}
\begin{align}
\varDelta\bm{j}_M =& \bm{j}_M - \left.\bm{j}_M\right|_{\Delta=0}\notag\\
=& \nabla\times\varDelta\bm{M}_{\mathrm{para}},
\end{align}
\begin{align}
\varDelta\bm{M}_{\mathrm{para}} =& -\frac{T}{V}\sum_{\bm{k},\omega_n,\alpha,\beta}e^{i\omega_n 0^{+}}\mu_s\bm{\sigma}_{\alpha\beta}\varDelta G_{\beta\alpha}(\bm{r};\bm{k},\omega_n)\notag\\
=& -i\mu_s 2\pi T\sum_{\omega_n>0,a}N_a\left< (-1)^{a+1}\hat{\bm{g}}_{\bm{k}}(g_a + 1) \right>_{\mathrm{FS}}.
\end{align}
Here,
\begin{equation}
\varDelta G_{\beta\alpha}(\bm{r};\bm{k},\omega_n) = G_{\beta\alpha}(\bm{r};\bm{k},\omega_n) - \left.G_{\beta\alpha}(\bm{r};\bm{k},\omega_n)\right|_{\Delta=0}
\end{equation}
and we have used the fact that $\left. g_a\right|_{\Delta=0} = -1$ for $\omega_n>0$. 

Then, we define the expression 
\begin{widetext}
\begin{align}
\Omega[\bm{A}, \Delta, f, \bar{f}] =& -\int d^3\bm{r}\left( \varDelta\bm{j}_S\cdot\bm{A} + \varDelta\bm{M}_{\mathrm{para}}\cdot\bm{B} \right) \notag\\
&\!\!\!\!\!\!\!\!\!\!\!\!\!\!\!\!\!\!\!\!\!\!\!\!\!\!\!\!\!\! + \int d^3\bm{r} \left[N|\Delta|^2\ln{\frac{T}{T_c}} + 2\pi T\sum_{\omega_n>0}\frac{N_a}{2}\left<\left(i\Delta^{\ast}w_a^{\ast}f_a + iw_a\Delta\bar{f}_a + \frac{|w_a\Delta|^2}{\omega_n} + (g_a+1)\left(2\omega_n + \frac{1}{2}\bm{v}_{\mathrm{F}}\cdot\nabla \ln\frac{f_a}{\bar{f}_a}\right)\right)\right>_{\mathrm{FS}}\right]
\label{frefl1}
\end{align}
\end{widetext}
as the functional from which the Eilenberger equations (\ref{eq:eilenberger_1}) and (\ref{eq:eilenberger_2}), the gap equation (\ref{eq:gap_equation}), and the difference of the current $\varDelta\bm{j}_S + \varDelta\bm{j}_M$ follow 
after variations with respect to $\bar{f}_a$ and $f_a$, $\Delta^{\ast}$, and $\bm{A}$, respectively.
Next, by replacing $f_a$ and $\bar{f}_a$ with the solutions $f_a[\Delta, \bm{A}]$ and $\bar{f}_a[\Delta, \bm{A}]$ of the Eilenberger equation under given $\Delta$ and $\bm{A}$, Eq.\,(\ref{frefl1}) is rewritten in the form 
\begin{widetext}
\begin{equation}
\widetilde{\Omega} = \Omega\left[\Delta, \bm{A}, f[\Delta, \bm{A}], \bar{f}[\Delta, \bm{A}]\right] = \int d^3\bm{r}\left[ N|\Delta|^2\ln{\frac{T}{T_c}} + 2\pi T\sum_{\omega_n>0,a}\frac{N_a}{2}\left<\left(\frac{|\Delta|^2}{\omega_n} - \frac{iw_a\Delta\bar{f}_a + i\Delta^{\ast}w_a^{\dag}f_a}{g_a-1}\right)\right>_{\mathrm{FS}} \right],
\end{equation}
\end{widetext}
obeying the conditions
\begin{equation}
\frac{\delta\widetilde{\Omega}}{\delta\Delta} = \left.\frac{\delta\Omega}{\delta\Delta}\right|_{\substack{f_a = f_a[\bm{\Delta},\bm{A}] \\ \bar{f}_a = \bar{f}_a[\bm{\Delta,\bm{A}}]}} = \frac{\delta F}{\delta\Delta},\\
\end{equation}
\begin{equation}
\frac{\delta\widetilde{\Omega}}{\delta\bm{A}} = \left.\frac{\delta\Omega}{\delta\bm{A}}\right|_{\substack{f_a = f_a[\bm{\Delta},\bm{A}] \\ \bar{f}_a = \bar{f}_a[\bm{\Delta,\bm{A}}]}} = \frac{\delta F}{\delta\bm{A}},\\
\end{equation}
\begin{equation}
\widetilde{\Omega}[\Delta=0, \bm{A}=\bm{0}] = F[\Delta=0, \bm{A}=\bm{0}] = 0.
\end{equation}
Thus, 
\begin{equation}
F = \widetilde{\Omega},
\end{equation}
which coincides with Eq.\,(\ref{eq:free_energy}).

\section{Calculation of $\Phi_a$}
Here, the relation (\ref{eq:Phi_lin_alg}) is derived.

For $\omega_n>0$, 
\begin{eqnarray}
\left[2 \left\{ \omega_n + i(-1)^{a+1}\mu_s\hat{\bm{g}}_{\bm{k}}\cdot\bm{B}\right\} + i\bm{v}_{\mathrm{F}}\cdot\bm{\Pi}\right]^{-1}&\notag\\
&\!\!\!\!\!\!\!\!\!\!\!\!\!\!\!\!\!\!\!\!\!\!\!\!\!\!\!\!\!\!\!\!\!\!\!\!\!\!\!\!\!\!\!\!\!\!\!\!\!\!\!\!\!\!\!\!\!\!\!\!\!\!\!\!\!\!\!\!\!\!\!\!\!\!\!\!\!\!\!\!\!\!\!\!\!\!\!\!\!\!\!\!\!\!\!\!\!\!\!\!\!\!\!\!\!\!\!\!\!\!
= \int_0^{\infty} d\rho e^{-2\left\{ \omega_n + i(-1)^{a+1}\mu_s\hat{\bm{g}}_{\bm{k}}\cdot\bm{B}\right\}\rho}e^{-i\bm{v}_{\mathrm{F}}\cdot\bm{\Pi}\rho}. \label{eq:inv_param_integ}
\end{eqnarray}
The operators
\begin{align}
a =& \frac{r_{H}}{\sqrt{2}}(\gamma^{-1/2}\Pi_{\bm{Q},z} - i\gamma^{1/2}\Pi_{\bm{Q},x}),\\
a^{\dag} =& \frac{r_{H}}{\sqrt{2}}(\gamma^{-1/2}\Pi_{\bm{Q},z} + i\gamma^{1/2}\Pi_{\bm{Q},x}),
\end{align}
which fulfill the relation $[a,a^{\dag}]=1$, are the annihilation and creation operators of the LLs (\ref{eq:LL}), where $\gamma = \xi_x/\xi_z$, $\bm{Q} = 2\delta N\bm{Q}_0$ ($\delta N = (N_2 - N_1)/(N_1 + N_2)$), $r_H = 1/\sqrt{2eH}$, and
\begin{equation}
\bm{\Pi}_{\bm{Q}} = \bm{\Pi} + \bm{Q}.
\end{equation}
With the identity $e^{A+B} = e^{[A,B]/2}e^{A}e^{B}$ in the case where $[A,[A,B]] = [B,[A,B]] = 0$, 
\begin{equation}
e^{-i\bm{v}_{\mathrm{F}}\cdot\bm{\Pi}_{\bm{Q}}\rho} = e^{-|s|^2\rho^2/2}e^{-is^{\ast}\rho a^{\dag}}e^{-is\rho a}.
\end{equation}
Thus, with the definition of the $N$th LL $\psi_N(\bm{r})$ (see Eq.\,(\ref{eq:LL}) in the main text),
\begin{equation}
\left[\psi_M^{\ast}e^{-i\rho\bm{v}_{\mathrm{F}}\cdot\bm{\Pi}_{\bm{Q}}}\psi_N \right]_{\mathrm{UC}} = e^{-|s|^2\rho^2/2}\mathcal{L}_{MN}(-is^{\ast}\rho), \label{eq:inner_prod_LL}
\end{equation}
where $[\ \cdot\ ]_{\mathrm{UC}}$ is the average over the unit cell and the relation
\begin{equation}
\left[\psi_M^{\ast} \psi_N \right]_{\mathrm{UC}} = \delta_{M,N}
\end{equation}
is used.
Here,
\begin{gather}
s = \frac{\gamma^{1/2}v_{\mathrm{F},z} + i\gamma^{-1/2}v_{\mathrm{F},x}}{\sqrt{2r_H^2}},\\
\mathcal{L}_{MN}(z) = \sum_{l=0}^{\min(M,N)}\frac{\sqrt{M!N!}}{(M-l)!(N-l)!l!}(z)^{M-l}(-z^{\ast})^{N-l}.
\end{gather}
From Eqs.\,(\ref{eq:inv_param_integ}) and (\ref{eq:inner_prod_LL}),
\begin{eqnarray}
[2\left\{ \omega_n + i(-1)^{a+1}\mu_s\hat{\bm{g}}_{\bm{k}}\cdot\bm{B}\right\} + i\bm{v}_{\mathrm{F}}\cdot\bm{\Pi}]^{-1}\Delta & \notag\\
&\!\!\!\!\!\!\!\!\!\!\!\!\!\!\!\!\!\!\!\!\!\!\!\!\!\!\!\!\!\!\!\!\!\!\!\!\!\!\!\!\!\!\!\!\!\!\!\!\!\!\!\!\!\!\!\!\!\!\!\!\!\!\!\!\!\!\!\!\!\!\!\!\!\!\!\!\!\!\!\!\!\!\!\!\!\!\!\!\!\!\!\!\!\!\!\!\!\!\!\!\!\!\!\!\!\!\!\!\!\!
= \int_0^{\infty} d\rho e^{-2\left\{ \omega_n + i(-1)^{a+1}\mu_s\hat{\bm{g}}_{\bm{k}}\cdot\bm{B}\right\}\rho}e^{i\bm{v}_{\mathrm{F}}\cdot\bm{Q}\rho}e^{-i\bm{v}_{\mathrm{F}}\cdot\bm{\Pi}_{\bm{Q}}\rho}\Delta \notag\\
&\!\!\!\!\!\!\!\!\!\!\!\!\!\!\!\!\!\!\!\!\!\!\!\!\!\!\!\!\!\!\!\!\!\!\!\!\!\!\!\!\!\!\!\!\!\!\!\!\!\!\!\!\!\!\!\!\!\!\!\!\!\!\!\!\!\!\!\!\!\!\!\!\!\!\!\!\!\!\!\!\!\!\!\!\!\!\!\!\!\!\!\!\!\!\!\!\!\!\!\!\!\!\!\!\!\!\!\!\!\!\!\!\!\!\!\!\!\!\!\!\!\!\!\!\!\!\!\!\!\!\!\!\!\!\!\!\!\!\!\!\!\!\!\!\!\!\!\!\!\!\!\!\!\!\!\!\!\!\!\!\!\!\!\!\!\!\!\!\!\!\!\!\!\!\!\!\!\!\!\!\!\!\!\!\!\!\!\!\!\!\!\!\!\!\!\!\!
= \psi_M\mathcal{M}^a_{MN}d_N
\end{eqnarray}
with the definition of the matrix 
\begin{align}
\mathcal{M}^a_{MN} &=\int_{0}^{\infty}d\rho e^{-(2\omega_n\rho + |s|^2\rho^2/2)}\notag\\
&\times e^{i\left\{\bm{v}_{\mathrm{F}}\cdot\bm{Q}-2(-1)^{a+1}\mu_s\hat{\bm{g}}_{\bm{k}}\cdot\bm{H}\right\}\rho}\mathcal{L}_{MN}(-is^{\ast}\rho),
\end{align}
which leads to Eq.\,(\ref{eq:Phi_lin_alg}).

\section{Derivation of LDOS}
Here, we define the retarded Green's function as usual:
\begin{eqnarray}
G^{\mathrm{R}}_{\alpha\beta}(\bm{r}_1,\bm{r}_2;t_1-t_2)&\notag\\
&\!\!\!\!\!\!\!\!\!\!\!\!\!\!\!\!\!\!\!\!\!\!\!\!\!\!\!\!\!\!\!\!\!\!\!\!\!\!\!\!\!\!\!\!\!\!\!\!\!\!= -i\Theta(t_1-t_2)\left<\left\{ c_{\alpha}(\bm{r}_1,t_1), c_{\beta}^{\dag}(\bm{r}_2,t_2) \right\}\right>_{eq},
\end{eqnarray}
where $\left\{\hat{A},\hat{B}\right\}=\hat{A}\hat{B}+\hat{B}\hat{A}$ is the anticommutator of arbitrary operators $\hat{A}$ and $\hat{B}$,
\begin{equation}
\Theta(t) = 
\begin{cases}
1\ &(t>0) \\
0\  &(t<0)
\end{cases}
\end{equation}
is the step function, and
\begin{equation}
\hat{X}(t) = e^{it(\mathcal{H}-\mu\mathcal{N})}\hat{X}e^{-it(\mathcal{H}-\mu\mathcal{N})}
\end{equation}
is the Heisenberg representation of any operator $\hat{X}$.
Its Wigner representation is defined as
\begin{align}
G^{\mathrm{R}}_{\alpha\beta}(\bm{r};\bm{k},E) &= \int d^3\bm{r}^{\prime}\, e^{-i\bm{k}\cdot\bm{r}^{\prime}}\int_{-\infty}^{+\infty}dt\, e^{iEt}\notag\\
&\times G^{\mathrm{R}}_{\alpha\beta}(\bm{r}+\bm{r}^{\prime}/2, \bm{r}-\bm{r}^{\prime}/2; t).
\end{align} 
As is well known, this can be obtained from $G_{\alpha\beta}(\bm{r};\bm{k},\omega_n)$ by the analytic continuation $i\omega_n\rightarrow E + i\eta$ ($\eta$ is an infinitesimal positive number).

The LDOS $N(\bm{r};E)$ can be defined by using this function.
\begin{equation}
N(\bm{r};E) = -\frac{1}{\pi}\frac{1}{V}\sum_{\bm{k},\alpha}\mathrm{Im}\,{G^{\mathrm{R}}_{\alpha\alpha}}(\bm{r}; \bm{k}, E).
\end{equation}
Finally, after performing the unitary transformation Eq.\,(\ref{eq:green_func_trans}) in the above expression, Eq.\,(\ref{eq:LDOS}) is obtained.

\end{document}